\documentclass[a4paper,11pt]{article}

\usepackage[utf8]{inputenc}
\usepackage[T1]{fontenc}

\usepackage{lmodern}
\usepackage{a4wide}
\usepackage{amsmath,amssymb,amsfonts}
\usepackage{color}
\usepackage[ruled,vlined,lines numbered]{algorithm2e}
\usepackage{hyperref}
\usepackage{tikz}
\usetikzlibrary{decorations.pathreplacing}
\usetikzlibrary{matrix}
\usetikzlibrary{patterns}
\usepackage{graphicx}
\usepackage[geometry]{ifsym}
\usepackage[caption=false]{subfig}
\usepackage{multirow}

\newcommand*\samethanks[1][\value{footnote}]{\footnotemark[#1]}

\newcommand{\Real}{\mathbb{R}}
\newcommand{\Complex}{\mathbb{C}}

\newcommand{\tr}{\ast}

\newcommand{\nice}[1]{\mathcal{#1}}

\newcommand{\bb}{\mathbf{b}}
\newcommand{\cc}{\mathbf{c}}

\newcommand{\redvel}{\nice{O}}

\newcommand{\mb}[1]{\left[\begin{array}{#1}}
\newcommand{\me}{\end{array}\right]}
\newcommand{\smb}{\left[\begin{smallmatrix}}
\newcommand{\sme}{\end{smallmatrix}\right]}

\newcommand{\SLICOT}{\textsf{SLICOT}~}

\newcommand{\alg}[1]{\textsf{#1}}
\newcommand{\tab}{\hspace*{0.5cm}}

\begin{document}

\title{Parallel solver for shifted systems in a hybrid CPU--GPU framework\thanks{This research has been supported by the Croatian Science Foundation under the grant HRZZ-9345.}}
\author{Nela Bosner\thanks{Faculty of Science, Department of Mathematics, University of Zagreb, Croatia.} \and Zvonimir Bujanovi\'{c}\samethanks \and Zlatko Drma\v{c}\samethanks}


\date{\today}
\maketitle

\begin{abstract}
	This paper proposes a combination of a hybrid CPU--GPU and a pure GPU software implementation of a direct algorithm for solving shifted linear systems $(A - \sigma I)X = B$ with large number of complex shifts $\sigma$ and multiple right-hand sides.
	Such problems often appear e.g. in control theory when evaluating the transfer function, or as a part of an algorithm performing
	interpolatory model reduction, as well as when computing pseudospectra and structured pseudospectra, or solving large linear systems of ordinary differential equations.
	The proposed algorithm first  jointly reduces the general full $n\times n$ matrix $A$ and the $n\times m$ full right-hand side matrix $B$ to the controller Hessenberg canonical form that facilitates efficient solution: $A$ is transformed to a so-called $m$-Hessenberg form and $B$ is made upper-triangular.
	This is implemented as blocked highly parallel CPU--GPU hybrid algorithm; individual blocks are reduced by the CPU, and the necessary updates of the rest of the matrix are split among the cores of the CPU and the GPU. To enhance parallelization, the reduction and the updates are overlapped.
	In the next phase, the reduced $m$-Hessenberg--triangular systems are  solved entirely on the GPU, with shifts divided into batches.
The benefits of such load distribution are demonstrated by numerical experiments. In particular, we show that our proposed implementation provides an excellent  basis for efficient implementations of computational methods in systems and control theory,  from evaluation of transfer function to the interpolatory model reduction.
%
	%
	%


\end{abstract}
%




\section{Introduction and preliminaries}
\label{intro}

The problem of solving shifted linear systems of equations
$
(A-\sigma I)x=\bb
$
arises in a variety of applications. The $n\times n$ matrix $A$ may or may not be structured (e.g.~sparse), and the real or complex shift $\sigma$ may take from a handful to several thousands of values. The right hand side $b$ may be constant or also shift dependent, $\bb=\bb(\sigma)$, and it may have several columns.
For instance, for a function $f$, $f(A)\bb$ can be efficiently approximated by numerically evaluated Cauchy integral
\begin{equation}
f(A)\bb = \frac{1}{2\pi\mathbf{i}}\int_{\Gamma}f(z)(zI-A)^{-1}\bb dz
\end{equation}
over a closed contour $\Gamma$ that encloses the spectrum of $A$, see e.g. \cite{HaleHighamTrefethen}. In this case $(\sigma_j I - A)^{-1}\bb$ is needed for the values $\sigma_j\in\Gamma$ required in a particular quadrature formula.

Computing the resolvent $(zI-A)^{-1}$ is also at the core of estimating the $\varepsilon$--pseudospectrum
\begin{equation}\label{eq:pseudospectrum}
\Lambda_{\varepsilon}(A) = \{ z\in \mathbb{C}:\| (zI-A)^{-1}\|_2 \ge \varepsilon^{-1}\},
\end{equation}
which requires estimates of $\| (zI-A)^{-1}\|_2$ over a grid of discrete points $z_j$. (Here the norm $\|\cdot\|_2$ is the spectral norm.)

If the dimension $n$ of the coefficient matrix $A$ is sufficiently large (e.g.~$n>10^4$) and the shift $\sigma$ takes e.g.~several hundreds up to few thousands values, then the mathematically simple task becomes a computationally challenging bottleneck in many applications. Hence, any effort to devise an efficient algorithm for solving a sequence of shifted linear systems on modern computers is justified, and any advance in software development immediately improves software packages whose efficiency depends on solving shifted systems. In this work, we offer such an improvement and illustrate its superior performance in an application.

Our motivation for this work has been mainly driven by applications in computational control, ranging from simple evaluation of matrix rational functions to model order reduction.  To set the stage, we briefly review some applications that require solving shifted linear systems.

Consider the LTI dynamical system
\begin{equation}\label{eq:LTI}
\dot{x}(t) = A x(t) + Bu(t), \;\; y(t)=Cx(t)
\end{equation}
with the system matrix $A \in \Real^{n \times n}$, the input  $B \in \Real^{n \times m}$, and the output matrix $C \in \Real^{p \times n}$, where $m, p\ll n$. Even the simple task of graphing the frequency response (Bode plot), i.e.~the values of the transfer function
\begin{align}
\label{intro:eq:TF}
\nice{G}(s) = C (sI-A)^{-1} B,\;\;s\in\mathbf{i}\Real\equiv \{ \mathbf{i}\omega\; :\; \omega\in\Real  \},
\end{align}
requires evaluations of $(A - \sigma_j I)^{-1} B$ for many values $\sigma_j = \mathbf{i}\omega_j\in\mathbf{i}\Real$. Already for moderately large $n,m,p$, this mere function evaluation may take annoyingly long time.

Computation similar to (\ref{intro:eq:TF}) is required in estimating the structured $\varepsilon$--pseudospectrum \cite{HinrichsenKelb}
\begin{equation}\label{eq:structured-pseudospectrum}
\Lambda_{\varepsilon}(A,B,C)=\{ z\in \mathbb{C}:\| C(zI-A)^{-1}B\|_2 \ge \varepsilon^{-1}\},
\end{equation}
i.e.~the set of complex numbers $z$ lying in the spectrum of $A + B E C$ for some $m\times p$ perturbation $E$ such that $\|E\|_2\leq\varepsilon$.

Another example where solving shifted systems consumes the major part of the computation time is interpolatory model order reduction. The task is to approximate (\ref{intro:eq:TF}) by a rational function $\nice{G}_r(s)$ of order $r$ with $r\ll n$.  One such method is the Iterative Rational Krylov Algorithm (IRKA) \cite{IRKA}, which finds an approximation $\nice{G}_r(s)$ that is locally optimal in the norm of the Hardy space $\mathcal{H}_2(\Complex_+)$. The local optimality of $\nice{G}_r(s)$ is a consequence of its Hermite interpolation property, i.e.~it tangentially interpolates $\nice{G}(s)$ and its derivative at particularly determined points -- mirror images of its own poles. The vicious circle is resolved by fixed point iterations.
The Hermite tangential interpolation at each iteration is achieved implicitly by the Petrov--Galerkin projection of (\ref{eq:LTI}) using the search and the test spaces as, respectively, column spaces of
\begin{eqnarray}
V_k &=& \begin{pmatrix} (A-\sigma_1^{(k)} I)^{-1} \bb_1^{(k)} &\ldots &  (A-\sigma_r^{(k)} I)^{-1} \bb_r^{(k)}\end{pmatrix},\;\;\; k = 1, 2, \ldots   \label{eq:IRKA:V}\\
W_k &=& \begin{pmatrix} (A-\sigma_1^{(k)} I)^{-T} \cc_1^{(k)} &\ldots &  (A-\sigma_r^{(k)} I)^{-T} \cc_r^{(k)}\end{pmatrix},\;\; k = 1, 2, \ldots   \label{eq:IRKA:W}
\end{eqnarray}
At each iteration $k$, the $r$ shifts $\sigma_i^{(k)}$ with the corresponding vectors $\bb_i^{(k)}$, $\cc_i^{(k)}$ are computed from the solution of a certain projected $r$ dimensional eigenvalue problem, and then $V_k$ and $W_k$ are computed. This computation is somewhat simplified in the SISO case ($m=p=1$) because then $\bb_i^{(k)}=B$, $\cc_i^{(k)}=C^T$ for all $i$, $k$. In the general case of many inputs and many outputs, the $\bb_i^{(k)}$'s and the $\cc_i^{(k)}$'s are confined to the column spaces of $B$ and $C^T$, respectively.

Hence, at each iteration IRKA requires solutions of $2r$ shifted linear systems of dimension $n$.
Take e.g.~$n=10^4$, $r=100$ and at least $20$ iterations -- the total is at least $4000$ shifted linear systems of dimension $10^4$. Moreover, since IRKA finds only a locally optimal approximation, and in the process of finding an appropriate reduced order model, a user may repeat the computation with different initial shifts $\sigma_i^{(0)}$ (hoping to find better local minimum of the approximation error), and perhaps with several values of $r$.
This design process may, altogether, require solutions of several thousands shifted linear systems. Similar situation is in the case of discrete systems \cite{BUNSEGERSTNER20101202}, with the appropriate version of IRKA. Further, methods such as the frequency domain POD \cite{kim-1998} also require solutions of shifted systems to compute the complex snapshots $(\mathbf{i}\omega_j I - A)^{-1}B$.

In a parametric model reduction $A=A(\wp)$, $B=B(\wp)$, $C=C(\wp)$ are parameter dependent, with the parameter $\wp\in\mathcal{P}\subset\mathbb{R}^d$, $d\geq 1$. Then a parametric formulation of IRKA \cite[Algorithm 4.1, Algorithm 5.1]{Baur-Beattie-Benner-Gugercin-2011} requires solutions of the systems (\ref{eq:IRKA:V}), $(\ref{eq:IRKA:W})$ over a parameter grid $\mathcal{P}_\sharp\subset\mathcal{P}$, so the total number of shifted systems to be solved is multiplied by the cardinality of $\mathcal{P}_\sharp$.

Another approach to parametrized model reduction is a hybrid method \cite{Baur-Benner-2009}, that uses balanced truncation to devise reduced order models for selected parameter values, and then combine them in an interpolatory fashion that again requires multiple evaluations of the transfer functions of the computed truncated systems. For more details and more examples of model reduction strategies that require multiple solutions of shifted systems we refer to the excellent survey \cite{Benner-Gugercin-Willcox-2013}.

{

\subsection{Solution methods}
If the matrix $A$ is large and sparse, such that the mapping $\Complex^n\ni v\mapsto A v \in\Complex^n$ is available as an efficient subroutine, then natural choices of methods for the computations (\ref{intro:eq:TF}), (\ref{eq:IRKA:V}, \ref{eq:IRKA:W}) are iterative methods based on Krylov subspaces, in particular because of their shift invariance. Successful examples are restarted GMRES \cite{Frommer-Glaessner-1998}, \cite{Zhong-Gu-2017}, BiCGStab \cite{Frommer-2003}, restarted FOM \cite{Simoncini-2003}.
Particularly tailored for the IRKA algorithm and the systems
(\ref{eq:IRKA:V}, \ref{eq:IRKA:W}) is the preconditioned BiCG \cite{Ahmad-Szyld-Gijzen-2017}.
In the applications in model reduction, the shifts used in (\ref{eq:IRKA:V}, \ref{eq:IRKA:W}) will slowly change and start to settle after some index $k$, allowing approximation based on previously computed subspaces, \cite{Recycl-Krylov-Parks-Sturler-2006}, \cite{Ahuja-Sturler-Gugercin-Chang-2012}, \cite{Ahuja-Benner-Sturler-Feng-2015}.

However, the performance of iterative methods often depends on the availability of good preconditioner (which may not be a simple task to ensure), and in some applications the matrix $A$ is not necessarily sparse, and its dimension $n$ is not extremely large, say $n$ is in tens of thousands at most. (An example of this are the hybrid methods for model reduction, discussed above.) In such situation, the total number of shifted systems to be solved, the required accuracy of the solution, and the computing platform (e.g.~massively parallel hardware, available optimized libraries) may motivate and justify  development of direct methods. Furthermore, even an iterative method for large sparse systems may need  direct solvers for the projected systems.

In the direct method paradigm, the most economic course of action is to reduce $A$ (and, if possible the right hand side) to a canonical form that is efficient to compute and convenient for solving many shifted linear systems. Typically, the initial decomposition is paid off after only a few shifts.
%
An example of such approach is the \SLICOT \cite{SLICOT}  routine \alg{TB05AD}, which first reduces the matrix $A$ to a Hessenberg form $\widetilde{A}$ via
an orthogonal transformation $Q$: $\widetilde{A} = Q^\tr A Q$. Then
$
	\nice{G}(\sigma_\ell) = -\widetilde{C} (\widetilde{A} - \sigma_\ell I)^{-1} \widetilde{B},
$
with $\widetilde{C} = C Q$ and $\widetilde{B} = Q^\tr B$. One now needs to solve linear systems with (shifted) Hessenberg matrices,
which is a much easier task.
Although the orthogonal transformation has to be done only once, the Hessenberg systems cannot be solved simultaneously for all shifts,
and are dealt with by processing only one shift at a time. This drastically reduces the potential for taking advantage of
the parallel architecture of contemporary computer systems.

The algorithms described in \cite{Beattie}, \cite{BosBD13} overcome this obstacle. In the first phase, the pair $(A, B)$ is
reduced to a so-called controller Hessenberg form: an orthogonal matrix $Q$ is constructed such that
$\widehat{A} = Q^\tr A Q$ is $m$-Hessenberg ($\widehat{A}_{i, j}=0$ for all $i > j+m$), and $\widehat{B} = Q^\tr B$ is upper triangular;
the matrix $\widehat{C} = CQ$ has no particular structure.
After this reduction, a carefully designed procedure is performed in order to compute
$
	\nice{G}(\sigma_\ell) = -\widehat{C} (\widehat{A} - \sigma_\ell I)^{-1} \widehat{B},
$
simultaneously for as many shifts as permitted by the computer memory capacity.
This whole procedure is well-suited for multicore platforms and makes good use of the cache memory structure:
its building blocks are predominantly BLAS3 operations, such as matrix multiplication, that involve large matrices.
Thus it performs much better than the implementation from \cite{SLICOT}.

In the context of computing the pseudospectra (\ref{eq:pseudospectrum}), (\ref{eq:structured-pseudospectrum}), one can e.g.~first reduce $A$ to (full or partial) Hessenberg or Schur form $F$, and then deploy an iterative method, such as the inverse iterations or the inverse Lanczos iterations, to compute the smallest singular value $\sigma_{\min}(z I-F)\equiv \| (z I - F)^{-1}\|_2^{-1}$,  see \cite{Lui}, \cite{Wright-Trefethen-2001}. These iterative methods require an efficient solver for shifted systems for several values of $z$; in some implementations the transposed system is also solved as in (\ref{eq:IRKA:V}), (\ref{eq:IRKA:W}), because the minimal singular value is computed as $\sigma_{\min}(z I-F) = \sqrt{\lambda_{\min}((zI-F)^*(z I-F))}$. With an efficient shifted Hessenberg solver, there is no need for reduction to the more expensive Schur form.  In  projection type methods, $A$ is accessed only as an operator/subroutine through matrix--vector products, and the resolvent is approximated from certain subspaces, see e.g.~\cite{BraconnierHigham}, \cite{TohTrefethen}, \cite{SimonciniGallopoulos}. If $B$ and $C^{*}$ are tall rectangular, full rank matrices of suitable dimensions, then the resolvent norm is approximated by $\| C(zI-A)^{-1}B\|_2$, where $C(zI-A)^{-1}B$ is the corresponding projected resolvent. The subspaces are usually generated by the Arnoldi or the Lanczos method, and e.g.~as in \cite{SimonciniGallopoulos} shifted Hessenberg systems are solved for many shifts.

\subsection{Our contributions in this work}
In this paper we adapt the algorithm of \cite{Beattie}, \cite{BosBD13} to a hybrid CPU--GPU setting, and offer detailed blueprints of an efficient software implementation for solving many shifted systems.
The initial controller Hessenberg form of $(A,B)$ is computed by a blocked highly parallel CPU--GPU hybrid algorithm; individual blocks are reduced by the CPU, and the necessary updates of the rest of the matrix are split among the cores of the CPU and the GPU; the reduction and the updates are overlapped.
In the second phase, the reduced shifted systems are solved entirely on the GPU. To enhance parallelism and data locality, the shifts are processed in batches, and the $m$ subdiagonals of the shifted system matrices are annihilated simultaneously for all shifts in a batch. 
This is the most demanding part of the computation and it requires RQ factorizations of many $m$-Hessenberg matrices; each factorization is computed by a different block of threads, while the updates mostly rely on cuBLAS routines.
In addition to the efficient BLAS3 operations, we make further use of independent operations that can be carried out simultaneously for different shifts. 
Besides providing a useful software tool for a variety of applications in scientific computing, we believe that our contribution can be also considered as an interesting and instructive case study for CPU+GPU matrix computation software development.

The rest of the  paper is organized as follows: in Section \ref{mhess} we describe an algorithm for the hybrid reduction of the matrix $A$
to the $m$-Hessenberg form. Section \ref{freq} contains the algorithm for solving shifted systems in the controller Hessenberg form,
including details such as parallel computation of the RQ factorization of a small matrix on the GPU, which serves as an
important auxiliary routine.
In \S \ref{subs:irka:syst}, we adapt our solver for large number of shifted linear systems to be suitable for the interpolatory model reduction algorithm IRKA \cite{IRKA}.
The final section shows the numerical experiments, comparing the implementation of \cite{BosBD13} that runs entirely on the CPU,
and the new hybrid implementation.
The performance of the GPU code over the CPU is shown in \S \ref{s:shifted-sys-results} to match the speedup of the DGEMM. 
In \S \ref{sec:irka:results}, we show that our code is a solid basis for an efficient implementation of IRKA.

\section{Reduction to controller Hessenberg form}
\label{mhess}
The key preprocessing step for efficient direct solution of many shifted system is reduction to a Hessenberg-type canonical form.
The matrix pair $(A,B)\in\Real^{n \times n}\times \Real^{n \times m}$ is reduced   to the controller--Hessenberg form.\footnote{For the sake of simplicity, we assume that, on input, $A$, $B$ and $C$ are real matrices. The adaptation to complex matrices is straightforward.}
If (depending on the computational task) we have the third matrix $C\in\Real^{p\times n}$, it will be updated accordingly, to preserve the equivalence with the original problem. The details are in Algorithm \ref{ALG:CHF}.



\begin{algorithm}[h]\label{ALG:CHF}
    \KwIn{$A\in\Real^{n \times n}$, $B\in\Real^{n \times m}$, $C\in\Real^{p \times n}$}
    \KwOut{$\widehat{A}\in\Real^{n \times n}$, $\widehat{B}\in\Real^{n \times m}$, $\widehat{C}\in\Real^{p \times n}$ in controller-Hessenberg form}

    \vspace*{0.3cm}
	Compute the QR-factorization $B = Q_B R_B$. Set $\widehat{B} = R_B$\;
	Apply the orthogonal similarity with $Q_B$, which is represented by Householder reflectors, onto $A$: $A \leftarrow Q_B^\tr A Q_B$.
		Transform $C \leftarrow C Q_B$\;
	Compute the $m$-Hessenberg form of the matrix $A$: $\widehat{A} = Q_A^\tr A Q_A$\;
	Transform $\widehat{C} = C Q_A$\;

	\caption{Transforming $(A,B,C)$ to the controller--Hessenberg form}\label{mhess:alg:contr_Hess}
\end{algorithm}
\noindent To implement this transformation on a hybrid CPU--GPU architecture, we adapt the CPU algorithm from \cite{BosBD13} by
using techniques similar to those described in \cite{Tomov}, where a hybrid (1-)Hessenberg reduction algorithm was proposed.

In terms of $\widehat{A}$, $\widehat{B}$, $\widehat{C}$, the computational tasks outlined in \S \ref{intro} allow better use of high performance hardware.
So, for instance, the evaluation of (\ref{intro:eq:TF}) is reformulated as
$
	\nice{G}(\sigma_\ell) = C(\sigma_\ell I - A)^{-1} B = - \widehat{C} (\widehat{A} - \sigma_\ell I)^{-1} \widehat{B},
$
where $\widehat{A}$ is $m$-Hessenberg, and $\widehat{B}$ is upper triangular. This form allows efficient CPU--GPU implementation of the computation $(\widehat{A} - \sigma_\ell I)^{-1} \widehat{B}$; we give the details in the next section. Note that in this case the orthogonal matrices $Q_B$ and $Q_A$ are canceled out and are not needed.

Let us now explain how this reduction simplifies the computation in the IRKA algorithm \cite{IRKA}. Set $Q = Q_B Q_A$.
In (\ref{eq:IRKA:V}), the vector $\bb_j^{(k)}$ is of the form $B \widehat{\bb}_j^{(k)}$ with certain $m\times 1$ vector $\widehat{\bb}_j^{(k)}$ (see \cite[\S 3.1]{Antoulas2010}). Hence, for the $j$th column of $V_k$, we have
$$
V_k(:,j)= (A-\sigma_j^{(k)} I)^{-1} \bb_j^{(k)} = (A-\sigma_j^{(k)} I)^{-1} B \widehat{\bb}_j^{(k)} = Q (\widehat{A} - \sigma_j I)^{-1} \widehat{B} \widehat{\bb}_j^{(k)} \equiv Q \widehat{V}_k(:,j) .
$$
Analogously, since $\cc_j^{(k)}= C^T \widehat{\cc}_j^{(k)}$, we have
$$
W_k(:,j)= (A-\sigma_j^{(k)} I)^{-T} \cc_j^{(k)} = (A-\sigma_j^{(k)} I)^{-T} C^T \widehat{\cc}_j^{(k)} = Q (\widehat{A} - \sigma_j I)^{-T} \widehat{C}^T \widehat{\cc}_j^{(k)} \equiv Q \widehat{W}_k(:,j) .
$$
%
Since in the next step IRKA computes the eigenvalues and the left and right eigenvectors of $W_k^T A V_k - \lambda W_k^T V_k$, in terms of new variables we have
$$
W_k^T A V_k - \lambda W_k^T V_k = \widehat{W}_k^T Q^T A Q \widehat{V}_k - \lambda \widehat{W}_k^T Q^T Q \widehat{V}_k = \widehat{W}_k^T \widehat{A} \widehat{V}_k - \lambda \widehat{W}_k^T  \widehat{V}_k .
$$
In an implementation of IRKA, one orthogonalizes the columns of $V_k$ and $W_k$ for better numerical stability. In the new representation, since $Q$ is orthogonal, this is equivalent to orthogonalizing the columns of $\widehat{V}_k$ and $\widehat{W}_k$.

As in the case of computing the transfer function (\ref{intro:eq:TF}), the reduction to controller--Hessenberg form is just a change of coordinates in the time domain representation that is an invariant for the computation in the frequency domain. Hence, one can equivalently run the IRKA algorithm in the new state space representation defined by $\widehat{A}$, $\widehat{B}$, $\widehat{C}$.


Going back to Algorithm \ref{ALG:CHF}, we note that
with the usual assumption $m, p \ll n$, the vast majority of the work is done in
line 3 of the algorithm;
this is the only step that does {$\redvel(n^3)$}
floating point operations on a matrix of order $n \times n$. Therefore, this critical step will be computed by the hybrid algorithm, while the other three steps
are done entirely on the CPU in order to minimize memory traffic between the CPU and the GPU.

The $m$-Hessenberg reduction is computed by using block Householder reflectors with two levels of blocking.
First, the matrix $A$ is divided into blocks (``panels''), each (except maybe the last one) containing $b$ consecutive columns.
Each block is further divided into so-called ``mini-blocks'' that have $m$ consecutive columns each.
Figure \ref{mhess:fig:main_alg} show this blocking scheme applied to the matrix $A$, midway through the reduction.

\subsection{Preliminaries: block oriented reduction}
For completeness, we now briefly recall the reduction procedure, and refer the reader to \cite{BosBD13} for full details.

The transformation matrix $Q_A$ from Algorithm \ref{mhess:alg:contr_Hess} has the form $Q_A = Q^{(1)} Q^{(2)} \ldots Q^{(k)}$,
where $k$ is the number of blocks in the matrix $A$. The blocks are denoted as $A^{(1)}, \ldots, A^{(k)}$.
Each $Q^{(i)}$ is a block Householder reflector, having form $Q^{(i)} = H^{(i)}_1 H^{(i)}_2 \ldots H^{(i)}_{\ell_i} = I - V^{(i)} T^{(i)} (V^{(i)})^\tr$,
where $\ell_i$ is the number of columns in the $i$-th block.
The matrices $H^{(i)}_j = I - \tau^{(i)}_j v^{(i)}_j (v^{(i)}_j)^\tr$ are Householder reflectors, computed so that zeros are introduced below the $m$-th
subdiagonal of the current block's $j$-th column.

After computing $H^{(i)}_j$, the matrix $Q^{(i)}$ is updated:
$Q_j^{(i)} = Q_{j-1}^{(i)} H^{(i)}_j = I - V_j^{(i)} T_j^{(i)} (V_j^{(i)})^\tr$ via
\begin{align}
    \label{mhess:eq:blockupdate_V}
    V_j^{(i)} =& \mb{cc} V_{j-1}^{(i)} & v^{(i)}_j \me; \\
    \label{mhess:eq:blockupdate_T}
    T_j^{(i)} =& \mb{cc}
                    T_{j-1}^{(i)} & T_{ij} \\ 0 & \tau^{(i)}_j
                 \me, \;
                 T_{ij} = -\tau^{(i)}_j T_{j-1}^{(i)} (V_{j-1}^{(i)})^\tr v^{(i)}_j .
\end{align}
Once the entire block $A^{(i)}$ is transformed to the $m$-Hessenberg form,
the remainder of the matrix $A$, i.e.~the blocks that are yet to be processed, has to be updated with the
newly computed block reflector:
\begin{align*}
    A \leftarrow (Q^{(i)})^\tr A Q^{(i)}
        &=  (I - V^{(i)} (T^{(i)})^\tr (V^{(i)})^\tr) A (I - V^{(i)} T^{(i)} (V^{(i)})^\tr) \\
        &= (I - V^{(i)} (T^{(i)})^\tr (V^{(i)})^\tr) (A - Y^{(i)} (V^{(i)})^\tr),
\end{align*}
where $Y^{(i)} = A V^{(i)} T^{(i)}$. Computing the second factor is called ``update from the right'', while computing the product with
the first factor is called ``update from the left''; it is irrelevant which one is executed first. 
Note that, during the block processing, one should also update the remainder of the current block with the partial block reflector $Q_j^{(i)}$
once $H_j^{(i)}$ is computed.
However, due to the structure of the partial block reflector, only updates from the left have to be performed for each column of the block, while
updates from the right have to be done only for every $m$-th column.
This is precisely the reason for introducing the mini-blocks.

The auxiliary matrix $Y^{(i)}$, which is also stored and maintained, thus only has to be updated for every $m$-th column of the block by
\begin{equation}
    \label{mhess:eq:update_Y}
    Y_{j-1+m}^{(i)} = \left[ \begin{array}{cc} Y_{j-1}^{(i)} & (-Y_{j-1}^{(i)} (V_{j-1}^{(i)})^\tr \nice{V}_j^{(i)} + A\nice{V}_j^{(i)})\nice{T}_j^{(i)} \end{array}\right] .
\end{equation}
Here $\nice{V}_j^{(i)}$ and $\nice{T}_j^{(i)}$ represent the mini-block reflector $\nice{Q}_j^{(i)} = I - \nice{V}_j^{(i)} \nice{T}_j^{(i)} (\nice{V}_j^{(i)})^\tr$,
which converts the mini-block of columns $j:j-1+m$ to the $m$-Hessenberg form, i.e.
$$
    V_{j-1+m}^{(i)} = \mb{cc} V_{j-1}^{(i)} & \nice{V}_j^{(i)} \me; \quad
    T_{j-1+m}^{(i)} = \mb{cc} T_{j-1}^{(i)} & \tilde{T}_{ij} \\ 0 & \nice{T}_j^{(i)} \me.
$$
Algorithm \ref{mhess:alg:mini_block} summarizes the processing of a single block, while Algorithm \ref{mhess:alg:outer} shows the
outer loop of the algorithm. To simplify notation, in Algorithm \ref{mhess:alg:mini_block} we have dropped the block index of all matrices
except $A^{(i)}$.

\begin{figure}[t]
    \begin{center}
        \begin{tikzpicture}[xscale=4, yscale=4]
            \draw [green, fill=green] (0.1, 1) -- (1, 1) -- (1, 0) -- (0.4, 0) -- (0.4, 0.5) -- (0, 0.9) -- (0,1) -- (0.1, 1);
            \draw [black] (0.1, 1) -- (1, 1) -- (1,0) -- (0.4, 0);
            \draw [black] (0.4, 0.5) -- (0, 0.9) -- (0,1) -- (0.1, 1);
            \draw [black, dashed] (0.4, -0.02) -- (0.4, 1.02);
            \draw [black, dashed] (0.5, -0.02) -- (0.5, 1.02);

            \draw [black, dashed] (0.3, 0.6) -- (1, 0.6);



            \draw [decorate,decoration={brace,amplitude=2pt},xshift=-0.02cm] (0,0.9) -- (0,1) node[black,midway,xshift=-0.6cm]{\footnotesize $m+1$};

            \draw [decorate,decoration={brace,amplitude=2pt, mirror},xshift=0.02cm] (1,0.6) -- (1,1) node[black,midway,xshift=0.2cm]{\footnotesize $k$};

            \draw [decorate,decoration={brace,amplitude=2pt,mirror},yshift=-0.04cm] (0.3,0) -- (0.4,0) node[black,midway,yshift=-0.2cm]{\footnotesize $m$};
            \draw [decorate,decoration={brace,amplitude=2pt,mirror},yshift=-0.04cm] (0.4,0) -- (0.5,0) node[black,midway,yshift=-0.2cm]{\footnotesize $m$};
            \draw [decorate,decoration={brace,amplitude=2pt,mirror},yshift=-0.04cm] (0.5,0) -- (0.6,0) node[black,midway,yshift=-0.2cm]{\footnotesize $m$};

            \draw [decorate,decoration={brace,amplitude=2pt},yshift=0.04cm] (0.3,1) -- (0.6,1) node[black,midway,yshift=0.25cm]{\footnotesize $b$};

            \draw [fill=black, fill opacity=0.3] (0.3, -0.02) rectangle (0.6, 1.02);
        \end{tikzpicture}
    \end{center}
    \caption{Reduction of the matrix $A$ to controller-Hessenberg form, with two levels of blocking: each block consisting of $b$ columns
        is further split into mini-blocks of $m$ columns. In the shaded block, the first mini-block is
        already transformed to the $m$-Hessenberg form.}
    \label{mhess:fig:main_alg}
\end{figure}
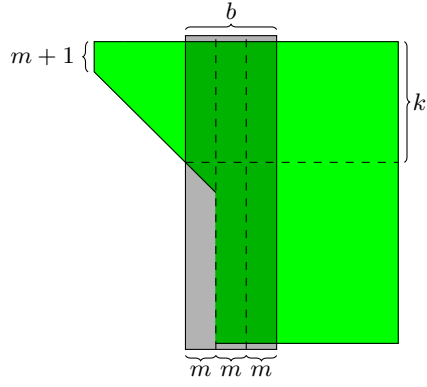

\begin{algorithm}
    \KwIn{block $A^{(i)}$, block size $b$, offset index $k$}
    \KwOut{partially updated block $A^{(i)}$, transformation matrices $V$, $T$, $Y$}

    \vspace*{0.3cm}
    \For{$j=1, 2, \ldots b$}
    {
        \If{$j > 1$}
        {
            Update $A^{(i)}(k+1:n, j)$ only from the left by applying $I-V_{j-1}T_{j-1}^\tr V_{j-1}^\tr$\;
        }

        Generate the elementary reflector $H_j$ to annihilate $A^{(i)}(k+j+1:n, j)$\;
        Compute $T_j(:, j)$ by using \eqref{mhess:eq:blockupdate_T}\;

        \If{$j$ mod $m=0$ or $j=b$}
        {
            $cMini = m \; or \; (b \; mod \; m)$; \tcp{current miniblock size}
            $nMini = \min\{b-j, m\}$; \tcp{next miniblock size}
            Compute  $Y_j(k+1:n, j-cMini+1:j)$ by using \eqref{mhess:eq:update_Y}\;
            Update the entire next miniblock from the right:\
                $A^{(i)}(k+1:n, j+1:j+nMini) = $ $A^{(i)}(k+1:n, j+1:j+nMini) - (Y_{j}V_{j}^\tr)(k+1:n, j+1:j+nMini)$\;
        }
    }
    Compute $Y(1:k, 1:b)$\;

    \caption{Processing of a block to reduce it to $m$-Hessenberg form}\label{mhess:alg:mini_block}
\end{algorithm}%

\begin{algorithm}
    \KwIn{$A\in\mathbb{R}^{n\times n}$, block size $b$, bandwidth $m$}
    \KwOut{$A$ converted to $m$-Hessenberg form}

    \vspace*{0.3cm}
    \For{$z=1, \; 1+b, \; 1+2\cdot b, \; 1+3\cdot b, \; \ldots$}
    {
        $i = (z-1)/b + 1$\;
        Process block $A^{(i)} = A(1:n, z:z+b-1)$ with $k=z+m-1$ by Algorithm \ref{mhess:alg:mini_block}
        to obtain $V$, $T$, $Y$\;

        \vspace*{0.3cm}
        \tcp{Apply block reflector from the right:}
        Call \texttt{xGEMM} to compute \
            $A(1:n, z+b:n) = A(1:n, z+b:n) - Y \cdot \left( V(k+b-m+1:n, 1:b) \right)^\tr$\;

        \vspace*{0.3cm}
        Call \texttt{xTRMM} to compute \
        $A(1:k, k+1:z+b-1) = A(1:k, k+1:z+b-1) - $\
            \hspace*{1cm}$Y(1:k, 1:b-m) \cdot \left( V(k+1:k+b-m, 1:b-m) \right)^\tr$\;

        \vspace*{0.3cm}
        \tcp{Apply block reflector from the left:}
        Call \texttt{xLARFB} to apply block reflector $(V, T)$ from left to $A( k+1:n, z+b:n )$\;
    }

    \caption{Block algorithm for reduction to the $m$-Hessenberg form}\label{mhess:alg:outer}
\end{algorithm}%

\subsection{A hybrid CPU-GPU implementation}
We now turn our attention on how to build a hybrid variant of the algorithm.
First we profile the CPU bound algorithm in order to detect which subtasks require the most computing resources
(here $n=8000$ and $m=20$; all percentages are relative to the total running time):

\begin{center}
\begin{tabular}{|l|c|}
    \hline
    subtask                                                             & CPU time   \\ \hline\hline  
    Processing of a block (Algorithm \ref{mhess:alg:mini_block})        & 41.53\%    \\ 
    $\bullet$ Computing $A\nice{V}_m$ in Line 9                         & (28.34\%)  \\ 
    $\bullet$ Line 11: update of $Y(1:k, 1:b)$                          & (8.73\%)   \\ 
    Out-of-block update (Lines 4--6 of Algorithm \ref{mhess:alg:outer}) & 58.47\%    \\ 
    $\bullet$ Line 4 to update $A$ from the right                       & (21.51\%)  \\ 
    $\bullet$ Line 6 to update $A$ from the left                        & (36.61\%)  \\ \hline 
\end{tabular}
\end{center}
\

\noindent
As we can see, most of the CPU time is spent doing the four bulleted operations; all of the four consist
entirely of matrix-multiply routines---either with full, or with triangular matrices according to their structure.
We distribute these operations to both the CPU and the GPU, proportionally to their computing power, and in order to
minimize data movement:
\begin{itemize}
    \item[(a)] Computation of $A\nice{V}_m$, followed by the update of the
    	matrix $Y(k+1:n, :)$ by the formula \eqref{mhess:eq:update_Y} is done on the GPU; it reduces to
    	calls to cuBLAS \cite{nvidia-cublas} matrix-multiply routines.
    \item[(b)] Line 11 of Algorithm \ref{mhess:alg:mini_block} will be computed by the CPU.

    \item[(c)] Lines 4--6 of Algorithm \ref{mhess:alg:outer} will be split to minimize communication between the CPU and the GPU, and
        to put heavier load on the GPU:
        \begin{enumerate}
            \item Update of $A$ from the left involves only $A(k+1:n, :)$; this will be done on the GPU;
            \item Update of $A(k+1:n, :)$ from the right will be done on the GPU;
            \item Update of $A(1:k, :)$ from the right will be done on the CPU.
        \end{enumerate}
\end{itemize}

At the beginning of the algorithm, the matrix $A$ is copied to the matrix $dA$ on the GPU.
As a block is being annihilated, each column is first updated from the left and a reflector is computed; this takes place on the CPU.
Then the computed reflector $v_j$ is copied to the GPU where we keep them in a separate $n \times b$ matrix $dV$. The matrix $dV$ has zeros above the
$m$-th subdiagonal and ones on it, so that products with it can be computed by using a single \alg{xGEMM}.


Once all columns in the current mini-block are annihilated, we already have the matrix $V_j$ ready on both the GPU ($dV$)
and the CPU (stored below the $m$-th subdiagonal of $A$). To update the block reflector using \eqref{mhess:eq:blockupdate_T},
we first copy the last $m$ columns (those that belong to the current mini-block) of $T$ to $dT$ at the GPU.
Then the GPU can update rows $k+1:n$ of the matrix $dA \cdot dV \cdot dT$, which overwrites parts of $dA$ not needed any more.
Once computed, these elements are transferred to the matrix $Y$ on the CPU. The matrix $Y$ is used on the CPU to update the next mini-block
from the right.
The pseudocode for the block processing is shown as Algorithm \ref{mhess:alg:GPU_panel}.

Conversion of the outer loop of the algorithm is straightforward; see Algorithms \ref{mhess:alg:GPU_GPUupdate} and \ref{mhess:alg:GPU_CPUupdate},
and note that we have moved the computation of $Y(1:k, :)$ from block processing to the CPU update.
As the CPU and the GPU each perform disjoint parts of the original algorithm,
some data in the matrices $A/dA$, $dV$, $T/dT$ and $Y$
may not reflect the actual situation at some point of the algorithm.
For example, Figure \ref{mhess:fig:GPU_afterOnePass} shows matrices $A$ and $dA$ after one pass of the main loop (process block + CPU update + GPU update).
The green area represents valid and up-to-date data, and the red area represents data which is not.
Fortunately, it is necessary to synchronize only a small part of the matrix $A$ in order for the algorithm to continue:
$b$ rows of the matrix $dA$ have to be copied to the CPU (line 3 of Algorithm \ref{mhess:alg:GPU_GPUupdate}),
as well as $b$ columns that belong to the next block (line 1 of Algorithm \ref{mhess:alg:GPU_panel}).
Such organization of the data movement and the workload distribution between the CPU and the GPU was inspired by \cite{Tomov}.

\begin{figure}
    \begin{center}
        \begin{minipage}{.45\textwidth}
            \begin{tikzpicture}[xscale=4, yscale=4]
                \draw [green, fill=green] (0.1, 1) -- (1, 1) -- (1, 0.6) -- (0.3, 0.6) -- (0, 0.9) -- (0,1) -- (0.1, 1);
                \draw [red, fill=red] (0.3, 0) -- (0.3, 0.6) -- (1, 0.6) -- (1, 0) -- (0.3, 0);

                \draw [black] (0.3, 0.6) -- (0, 0.9) -- (0,1) -- (1, 1) -- (1, 0.6);
                \draw [black] (0.3, 0.6) -- (0.3, 0) -- (1,0) -- (1, 0.6);

                \draw [black, dashed] (0.3, 0.6) -- (1, 0.6);



                \draw [decorate,decoration={brace,amplitude=2pt},xshift=-0.02cm] (0,0.9) -- (0,1) node[black,midway,xshift=-0.6cm]{\footnotesize $m+1$};
                \draw [decorate,decoration={brace,amplitude=2pt, mirror},xshift=0.02cm] (1,0.6) -- (1,1) node[black,midway,xshift=0.2cm]{\footnotesize $k$};
                \draw [decorate,decoration={brace,amplitude=2pt},yshift=0.04cm] (0.3,1) -- (0.5,1) node[black,midway,yshift=0.25cm]{\footnotesize $b$};

                \draw [fill=black, fill opacity=0.3] (0.28, -0.02) rectangle (0.5, 1.02);
                \draw (0.5, -0.1) node{\textsf{CPU: $A$}};
            \end{tikzpicture}
        \end{minipage}\hfill%
        \begin{minipage}{.45\textwidth}
            \begin{tikzpicture}[xscale=4, yscale=4]
                \draw [red, fill=red] (0.1, 1) -- (1, 1) -- (1, 0.6) -- (0.3, 0.6) -- (0, 0.9) -- (0,1) -- (0.1, 1);
                \draw [green, fill=green] (0.3, 0) -- (0.3, 0.6) -- (1, 0.6) -- (1, 0) -- (0.3, 0);

                \draw [black] (0.3, 0.6) -- (0, 0.9) -- (0,1) -- (1, 1) -- (1, 0.6);
                \draw [black] (0.3, 0.6) -- (0.3, 0) -- (1,0) -- (1, 0.6);

                \draw [black, dashed] (0.3, 0.6) -- (1, 0.6);



                \draw [decorate,decoration={brace,amplitude=2pt},xshift=-0.02cm] (0,0.9) -- (0,1) node[black,midway,xshift=-0.6cm]{\footnotesize $m+1$};
                \draw [decorate,decoration={brace,amplitude=2pt, mirror},xshift=0.02cm] (1,0.6) -- (1,1) node[black,midway,xshift=0.2cm]{\footnotesize $k$};
                \draw [decorate,decoration={brace,amplitude=2pt},yshift=0.04cm] (0.3,1) -- (0.5,1) node[black,midway,yshift=0.25cm]{\footnotesize $b$};


                \draw [fill=black, fill opacity=0.3] (0.28, -0.02) rectangle (0.5, 1.02);

                \draw [pattern=north west lines] (0.3, 0.0) rectangle (0.5, 0.6);
                \draw [pattern=north east lines] (0.5, 0.4) rectangle (1.0, 0.6);
                \draw [decorate,decoration={brace,amplitude=2pt, mirror},xshift=0.02cm] (1,0.4) -- (1,0.6) node[black,midway,xshift=0.2cm]{\footnotesize $b$};

                \draw (0.5, -0.1) node{\textsf{GPU: $dA$}};
            \end{tikzpicture}
        \end{minipage}
    \end{center}

    \caption{Matrices $A$ (on the CPU) and $dA$ (on the GPU) after one pass through the outer loop of Algorithm \ref{mhess:alg:GPU_outer}.
        Green color indicates elements that are up-to-date on each device, red color indicates those that are not.
        Shaded is the next block to be processed. Note that both devices need only their ``green elements'' in order
        to continue with the algorithm, except for the two thin strips {(marked with the diagonal zebra lines)}
        that need to be copied from the GPU back to the CPU.}
    \label{mhess:fig:GPU_afterOnePass}
\end{figure}
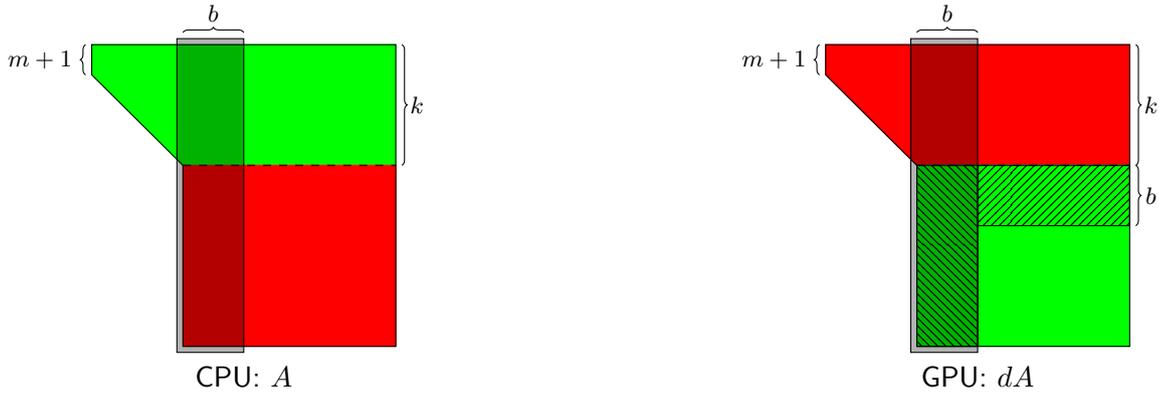


Another major observation is that the CPU update task can be done in parallel with the GPU update of the current block
and the processing of the next one.
To balance the amount of work,
only a few cores of the CPU will be dedicated to processing of a block, and all the others will be doing the CPU update.
The latter task is more time demanding for the CPU and the amount of work in it grows toward the end of the reduction. This is
why we further split the job of the CPU update among the cores: each core will get several consecutive
rows of $Y$ and $A$ to compute. This too can be done in parallel; should the number of cores be large enough, the computation can be
completed just before the GPU update of the current block and the block processing of the next one have both finished.
See Figure \ref{mhess:fig:GPU_parallelism}.

The timing breakdown now reads as follows (here $n=8000$ and $m=20$; $2$ cores are processing the block and $6$ CPU cores are computing the update):
\begin{center}
\begin{tabular}{|l|c|}
    \hline
    subtask                                                     & total time \\ \hline\hline 
    Processing of a block (Algorithm \ref{mhess:alg:GPU_panel}) & 51.21\%   \\ 
    $\bullet$ Computing $A\nice{V}_m$                           & (33.48\%) \\ \hline 
    GPU update (Algorithm \ref{mhess:alg:GPU_GPUupdate})        & 32.32\%   \\ 
    $\bullet$ Line 1 : update of $A(k+1,:)$ from the right      & (10.00\%) \\ 
    $\bullet$ Line 2 : update of $A(k+1,:)$ from the left       & (22.22\%) \\ \hline 
    CPU update (Algorithm \ref{mhess:alg:GPU_CPUupdate})        & 54.15\%   \\ 
    $\bullet$ Line 1 : update of $Y(1:k, 1:b)$                  & (27.59\%) \\ 
    $\bullet$ Line 2 : update of $A(1:k, z+b:n)$                & (26.56\%) \\ \hline 
\end{tabular}
\end{center}
\

\noindent
Note that the percentages don't add up to 100\% because the subtasks now overlap in time.

%
%

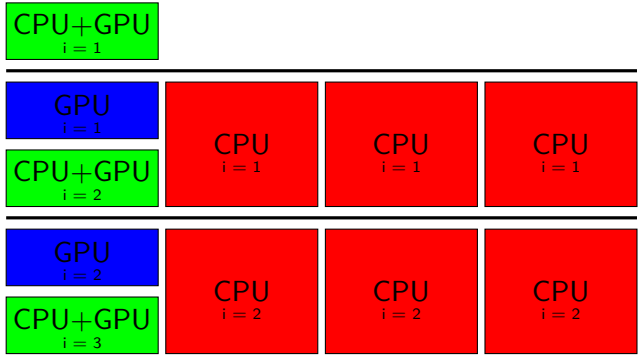
\begin{figure}

    \newcommand{\Full}[3]{%
        \begin{tikzpicture}{0}{0}{#3}{#2}
            \draw [black, fill=#1] (0,0) rectangle (#3,#2);
        \end{tikzpicture}}

    \newcommand{\Void}[2]{%
        \begin{tikzpicture}{0}{0}{#2}{#1}
            \draw [white] (0,0) rectangle (#2,#1);
        \end{tikzpicture}}

    {
        \centering

        \begin{minipage}{.6\textwidth}%
            \begin{tikzpicture}[xscale = 2, yscale = 3]
            \draw [fill=green] (0, 0.85) rectangle (1, 1.1);
            \draw (0.5, 1.0) node {\textsf{CPU+GPU}}; \draw (0.5, 0.9) node {\tiny{$\mathsf{i=1}$}};

            \draw [very thick] (0, 0.8) -- (4.15, 0.8);

            \draw [fill=blue] (0, 0.5) rectangle (1, 0.75);
            \draw (0.5, 0.65) node {\textsf{GPU}}; \draw (0.5, 0.55) node {\tiny{$\mathsf{i=1}$}};
            \draw [fill=green] (0, 0.2) rectangle (1, 0.45);
            \draw (0.5, 0.35) node {\textsf{CPU+GPU}};  \draw (0.5, 0.25) node {\tiny{$\mathsf{i=2}$}};

            \draw [fill=red] (1.05, 0.2) rectangle (2.05, 0.75);
            \draw (1.55, 0.475) node {\textsf{CPU}}; \draw (1.55, 0.375) node {\tiny{$\mathsf{i=1}$}};
            \draw [fill=red] (2.1, 0.2) rectangle (3.1, 0.75);
            \draw (2.6, 0.475) node {\textsf{CPU}}; \draw (2.6, 0.375) node {\tiny{$\mathsf{i=1}$}};
            \draw [fill=red] (3.15, 0.2) rectangle (4.15, 0.75);
            \draw (3.65, 0.475) node {\textsf{CPU}}; \draw (3.65, 0.375) node {\tiny{$\mathsf{i=1}$}};

            \draw [very thick] (0, 0.15) -- (4.15, 0.15);

            \draw [fill=blue] (0, -0.15) rectangle (1, 0.1);
            \draw (0.5, 0.0) node {\textsf{GPU}};  \draw (0.5, -0.1) node {\tiny{$\mathsf{i=2}$}};
            \draw [fill=green] (0, -0.45) rectangle (1, -0.2);
            \draw (0.5, -0.3) node {\textsf{CPU+GPU}}; \draw (0.5, -0.4) node {\tiny{$\mathsf{i=3}$}};

            \draw [fill=red] (1.05, -0.45) rectangle (2.05, 0.1);
            \draw (1.55, -0.175) node {\textsf{CPU}}; \draw (1.55, -0.275) node {\tiny{$\mathsf{i=2}$}};
            \draw [fill=red] (2.1, -0.45) rectangle (3.1, 0.1);
            \draw (2.6, -0.175) node {\textsf{CPU}}; \draw (2.6, -0.275) node {\tiny{$\mathsf{i=2}$}};
            \draw [fill=red] (3.15, -0.45) rectangle (4.15, 0.1);
            \draw (3.65, -0.175) node {\textsf{CPU}}; \draw (3.65, -0.275) node {\tiny{$\mathsf{i=2}$}};
        \end{tikzpicture}
        \end{minipage}%
        \begin{minipage}{.45\textwidth}
            \Full{green}{.3}{.3} hybrid processing of the blocks: \\
                \Void{.3}{.3} $Q^{(i)} = I - V^{(i)} T^{(i)} (V^{(i)})^\tr$, $i=1, 2, 3, \ldots$ \\[.3cm]
            \Full{blue}{.3}{.3} left updates $A \leftarrow (Q^{(i)})^\tr A$, GPU part \\
                \Void{.3}{.3} of the right updates $A \leftarrow A Q^{(i)}$ \\[.3cm]
            \Full{red}{.3}{.3} CPU part of the right updates \\
                \Void{.3}{.3} $A \leftarrow A Q^{(i)}$
        \end{minipage}
    }

    \caption{Parallelism in the hybrid algorithm for the $m$-Hessenberg reduction.
    In this figure, the block processing and the GPU--update is done on a single CPU core, while
    the CPU--update algorithm spreads across all other cores of the CPU.
    The thick black lines denote synchronization points.}
    \label{mhess:fig:GPU_parallelism}
\end{figure}

\begin{algorithm}
     \KwIn{block $A^{(i)}$, block size $b$, indices $k$ and $z$ with values as in line 4 of Algorithm \ref{mhess:alg:GPU_outer}}
     \KwOut{partially updated block $A^{(i)}$, transformation matrices $Y$ and $T$}

     \vspace*{0.3cm}
     Copy the current block $dA(k+1:n, z:z+b-1)$ from the GPU to $A^{(i)}(k+1:n, :)$ on the CPU\;
     \For{$j=1, 2, \ldots b$}
     {
         \If{$j > 1$}
         {
             Update $a_{j}=\text{the } j\text{-th column of } A^{(i)}$ on the CPU, only from the left side\;
         }

         Generate the elementary reflector to annihilate $a_{j}$ on the CPU and store it in $a_{j}(k+j:n)$\;
         Copy $a_{j}(k+j:n)$ to $dV(k+j:n,j)$ on the GPU\;

         Compute $T_j(:, j)$ on the CPU\;

         \vspace*{0.3cm}
         \If{$j$ mod $m=0$ or $j=b$}
         {
             $cMini = m \; or \; (b \; mod \; m)$; \tcp{current mini-block size}
             $nMini = \min\{b-j, m\}$; \tcp{next mini-block size}
             \vspace*{0.3cm}

             Copy last $cMini$ columns of $T_j$ from CPU to $dT_j$ on the GPU\;
             Compute  $dY_j(k+1:n, j-cMini+1:j)$ on the GPU\;
             Copy computed part of $dY_j$ to $Y_j$ on the CPU\;

             \vspace*{0.3cm}
             Update the entire next mini-block from the right on the CPU\;
         }
     }
     \caption{Hybrid processing of a block}\label{mhess:alg:GPU_panel}
 \end{algorithm}%

\begin{algorithm}
    Using $(dY, dV)$, apply the block reflector to update $dA(k+1:n, z+b:n)$ from the right on the GPU\;

    \vspace*{0.3cm}
    Using $(dV, dT)$, apply the block reflector to update $dA(k+1:n, z+b:n)$ from the left on the GPU\;

    \vspace*{0.3cm}
    Copy $dA(k+1:k+b, z+b:n)$ to $A$\;

    \caption{GPU update in the outer loop}\label{mhess:alg:GPU_GPUupdate}
\end{algorithm}%

\begin{algorithm}
    Compute $Y(1:k, 1:b)$\;

    \vspace*{0.3cm}
    Using $Y$, apply the block reflector to update $A(1:k, z+b:n)$ from the right on the CPU\;

    \caption{CPU update in the outer loop}\label{mhess:alg:GPU_CPUupdate}
\end{algorithm}%

\begin{algorithm}
    \KwIn{$n\times n$ matrix $A$, block size $b$, bandwidth $m$}
    \KwOut{$A$ converted to $m$-Hessenberg form}

    \vspace*{0.3cm}
    Copy the matrix $A$ from CPU to $dA$ on the GPU\;
    \For{$z=1, \; 1+b, \; 1+2\cdot b, \; 1+3\cdot b, \; \ldots$}
    {
        $i = (z-1)/b + 1$; $k = z+m-1$\;
        Process block $A^{(i)} = A(1:n, z:z+b-1)$ using hybrid Algorithm \ref{mhess:alg:GPU_panel} \;

        \vspace*{0.3cm}
        Call Algorithm \ref{mhess:alg:GPU_CPUupdate} asynchronously to compute $Y(1:k, :)$ and update $A(1:k, :)$ on the CPU from the right \;
        \vspace*{0.3cm}

        Call Algorithm \ref{mhess:alg:GPU_GPUupdate} to update $dA(k+1:n, :)$ on the GPU from both the left and the right \;
    }

    \caption{Hybrid algorithm for the $m$-Hessenberg reduction of $A$}
    \label{mhess:alg:GPU_outer}
\end{algorithm}%

\newpage

\section{Solving shifted systems in Hessenberg form}
\label{freq}
By virtue of the reduction described in \S \ref{mhess}, we henceforth assume that in (\ref{eq:LTI}), (\ref{intro:eq:TF}), (\ref{eq:structured-pseudospectrum}) the matrices $A\in \mathbb{R}^{n\times n}$, $B\in \mathbb{R}^{n\times m}$, and $C\in \mathbb{R}^{p\times n}$ are in the controller Hessenberg form, i.e.~having the following zero-patterns:

$$
A = \left[\begin{array}{c} \begin{tikzpicture}[xscale=1, yscale=1]
\draw [black, fill=green] (0,1) -- (1, 1) -- (1,0) -- (0,1);
\draw [black] (0.9,0) -- (0, 0.9);
\draw [black] (0.8,0) -- (0, 0.8);
\end{tikzpicture} \end{array}\right],
B = \left[\begin{array}{c} \begin{tikzpicture}[xscale=0.3, yscale=0.3]
\draw [black, fill=blue] (0,1) -- (1, 1) -- (1,0) -- (0,1);
\end{tikzpicture}\\ \begin{tikzpicture}{0}{0}{0.3}{0.7}
\draw [white] (0,0) rectangle (0.3,0.7);
\end{tikzpicture} \end{array}\right],
C = \left[\begin{array}{c} \begin{tikzpicture}{0}{0}{1}{0.3}
\draw [black, fill=red] (0,0) rectangle (1,0.3);
\end{tikzpicture} \end{array}\right] .
$$
The standard procedure for inverting $A-\sigma_\ell I$ is based on the RQ-factorization $A - \sigma_\ell I = R_\ell Q_\ell$ for every shift $\sigma_\ell$, where
$$
R_\ell = \left[\begin{array}{c} \begin{tikzpicture}[xscale=1, yscale=1]
\draw [black, fill=green] (0,1) -- (1, 1) -- (1,0) -- (0,1);
\end{tikzpicture} \end{array}\right],
Q_\ell = \left[\begin{array}{c} \begin{tikzpicture}[xscale=1, yscale=1]
\draw [black, fill=green] (0,1) -- (1, 1) -- (1,0) -- (0,1);
\draw [black] (0.9,0) -- (0, 0.9);
\draw [black] (0.8,0) -- (0, 0.8);
\end{tikzpicture} \end{array}\right].
$$
In this way, the evaluation of the transfer function reduces to $\nice{G}(\sigma_\ell) = (C Q_\ell^{*}) (R_\ell^{-1} B)$.
Due to the triangular form of $B$, {where each column of $B$ has at most $m$ nontrivial components}, the computation is further reduced to:
\begin{enumerate}
\item solving the $m\times m$ triangular system $Y_\ell = R_\ell(1:m,1:m)^{-1} B(1:m,1:m)$,
\item matrix multiplication $C_\ell = C Q_\ell^{*}$,
\item matrix multiplication $\nice{G}(\sigma_\ell) = C_\ell(:,1:m) \cdot Y_\ell$.
\end{enumerate}
Note that only the leading $m\times m$ block of $R_\ell$ is needed,
which further simplifies the computation of the RQ factorization (both in number of flops and memory management); also, only a part of the transformed $C$ is needed. As a result, for each $\sigma_\ell$ the computation is organized to compute only $R_\ell(1:m,1:m)$ and $C_\ell(:,1:m)$.

\subsection{Simultaneous handling of a batch of shifts}\label{subsect-seq-alg-tfe}
Our next goal is to show how one can simultaneously factorize $A - \sigma_\ell I = R_\ell Q_\ell$, for a batch of $s$ shifts. In the following series of figures, we illustrate that a large part of the data can be shared and reused when computing the RQ factorizations for several shifts at once. In the left half of the figures, we compute the factorization for a ``blue'' shift $\sigma_{\ell_1}$, and in the right half for a ``red'' shift $\sigma_{\ell_2}$. Green parts of the figures show the data which is the same for both shifts during the computation, while the blue and the red parts show the data that depends on the shift.

The RQ factorizations are performed in blocks of $n_{b}$ rows, starting from the bottom of the matrices, and moving towards the top.
Due to the $m$-Hessenberg form of the matrix $A-\sigma_\ell I$,
transformation of each block to the triangular form will affect $m+n_b$ consecutive columns of $A-\sigma_\ell I$ and $C$.
Since only a part of the RQ factorization is needed, a sliding window $Z^{(\ell )}$, which is an auxiliary array consisting of $(p+n)\times (m+n_{b})$ elements, is used to store $m+n_{b}$ columns of the thus far transformed matrices $C$ and $A-\sigma_\ell I$, organized as $\left[ \begin{smallmatrix} C\\ A-\sigma_\ell I \end{smallmatrix}\right]$.

\begin{tikzpicture}[xscale=3, yscale=3]
        \draw [black, fill=green] (0,1) -- (1, 1) -- (1,0) -- (0.7, 0) -- (0, 0.7) -- (0,1);
        \draw [blue, very thick] (0, 1) -- (1, 0);

        \draw [black, fill=green] (0, 1.02) rectangle (1, 1.2);
        \draw (-0.2, 0.5) node {$A - \textcolor{blue}{\sigma_{\ell_1}}I$};
        \draw (-0.2, 1.1) node {$C$};

        \draw [fill=black, fill opacity=0.3] (0.6, 0) rectangle (1, 1.2);
        \draw [black] (0.6, 0.1) -- (1, 0.1);
        \draw [->, blue] (0.45,0.05) -- (0.6,0.05);
        \draw (0.3,0.05) node {$\textcolor{blue}{Z_{block}^{(\ell_1)}}$};

        \draw (0.8, 1.3) node {$\textcolor{blue}{Z^{(\ell_1)}}$};

        \draw [decorate,decoration={brace,amplitude=2pt, mirror},xshift=0.22cm] (1,1.02) -- (1,1.2) node[black,midway,xshift=0.2cm]{\footnotesize $p$};
        \draw [decorate,decoration={brace,amplitude=2pt, mirror},xshift=0.22cm] (1,0) -- (1,1) node[black,midway,xshift=0.2cm]{\footnotesize $n$};
        \draw [decorate,decoration={brace,amplitude=2pt, mirror},xshift=0.02cm] (1,0) -- (1,0.1) node[black,midway,xshift=0.3cm]{\footnotesize $n_b$};
        \draw [decorate,decoration={brace,amplitude=2pt, mirror},yshift=-0.04cm] (0.6,0) -- (1,0) node[black,midway,yshift=-0.25cm]{\footnotesize $m+n_b$};

\begin{scope}[shift={(2.5,0)}]
        \draw [black, fill=green] (0,1) -- (1, 1) -- (1,0) -- (0.7, 0) -- (0, 0.7) -- (0,1);
        \draw [red, very thick] (0, 1) -- (1, 0);

        \draw [black, fill=green] (0, 1.02) rectangle (1, 1.2);
        \draw (-0.2, 0.5) node {$A - \textcolor{red}{\sigma_{\ell_2}}I$};
        \draw (-0.2, 1.1) node {$C$};

        \draw [fill=black, fill opacity=0.3] (0.6, 0) rectangle (1, 1.2);
        \draw [black] (0.6, 0.1) -- (1, 0.1);
        \draw [->, red] (0.45,0.05) -- (0.6,0.05);
        \draw (0.3,0.05) node {$\textcolor{red}{Z_{block}^{(\ell_2)}}$};

        \draw (0.8, 1.3) node {$\textcolor{red}{Z^{(\ell_2)}}$};

        \draw [decorate,decoration={brace,amplitude=2pt, mirror},xshift=0.22cm] (1,1.02) -- (1,1.2) node[black,midway,xshift=0.2cm]{\footnotesize $p$};
        \draw [decorate,decoration={brace,amplitude=2pt, mirror},xshift=0.22cm] (1,0) -- (1,1) node[black,midway,xshift=0.2cm]{\footnotesize $n$};
        \draw [decorate,decoration={brace,amplitude=2pt, mirror},xshift=0.02cm] (1,0) -- (1,0.1) node[black,midway,xshift=0.3cm]{\footnotesize $n_b$};
        \draw [decorate,decoration={brace,amplitude=2pt, mirror},yshift=-0.04cm] (0.6,0) -- (1,0) node[black,midway,yshift=-0.25cm]{\footnotesize $m+n_b$};
\end{scope}
\end{tikzpicture}

\noindent
\emph{Step 1:} For all $\ell =1,\ldots ,s$, we first compute the RQ-factorizations $Z_{block}^{(\ell )}=R_{block}^{(\ell )} \cdot P^{(\ell )}$ of the last $n_b$ rows of the current blocks $Z^{(\ell )}$, stored in  $Z_{block}^{(\ell )}$. All these factorizations are independent, and can be computed in parallel. We call this step ``small RQ-batched''.

\begin{tikzpicture}[xscale=3, yscale=3]
        \draw [black, fill=green] (0.1, 1) -- (1, 1) -- (1,0.1) -- (0.6, 0.1) -- (0, 0.7) -- (0,1) -- (0.1, 1);
        \draw [blue, very thick] (0, 1) -- (0.9, 0.1);

        \draw [black, fill=green] (0, 1.02) rectangle (1, 1.2);
        \draw (-0.2, 0.5) node {$A - \textcolor{blue}{\sigma_{\ell_1}}I$};
        \draw (-0.2, 1.1) node {$C$};

        \draw [fill=blue] (1, 0.1) -- (0.9, 0.1) -- (1, 0);

        \draw [fill=black, fill opacity=0.3] (0.6, 0) rectangle (1, 1.2);
        \draw [black] (0.6, 0.1) -- (1, 0.1);

        \draw (1, 0.05) node[anchor=west] {$\textcolor{blue}{R_{block}^{(\ell_1)}}$};

        \draw [fill=blue] (1.05, 0.3) rectangle (1.25, 0.5);
        \draw (1.01, 0.58) node[anchor=west] {$\textcolor{blue}{P^{(\ell_1)^*}}$};

        \draw (0.8, 1.3) node {$\textcolor{blue}{Z^{(\ell_1)}}$};
\begin{scope}[shift={(2.5,0)}]
        \draw [black, fill=green] (0.1, 1) -- (1, 1) -- (1,0.1) -- (0.6, 0.1) -- (0, 0.7) -- (0,1) -- (0.1, 1);
        \draw [red, very thick] (0, 1) -- (0.9, 0.1);

        \draw [black, fill=green] (0, 1.02) rectangle (1, 1.2);
        \draw (-0.2, 0.5) node {$A - \textcolor{red}{\sigma_{\ell_2}}I$};
        \draw (-0.2, 1.1) node {$C$};

        \draw [fill=red] (1, 0.1) -- (0.9, 0.1) -- (1, 0);

        \draw [fill=black, fill opacity=0.3] (0.6, 0) rectangle (1, 1.2);
        \draw [black] (0.6, 0.1) -- (1, 0.1);

        \draw (1, 0.05) node[anchor=west] {$\textcolor{red}{R_{block}^{(\ell_2)}}$};

        \draw [fill=red] (1.05, 0.3) rectangle (1.25, 0.5);
        \draw (1.01, 0.58) node[anchor=west] {$\textcolor{red}{P^{(\ell_2)^*}}$};

        \draw (0.8, 1.3) node {$\textcolor{red}{Z^{(\ell_2)}}$};
\end{scope}
\end{tikzpicture}

\noindent
\emph{Step 2:} The rest of $Z^{(\ell )}$ is updated by $P^{(\ell )}\in \mathbb{C}^{(n_{b}+m)\times (n_{b}+m)}$ from the right.

\begin{tikzpicture}[xscale=3, yscale=3]
        \draw [black, fill=green] (0.1, 1) -- (1, 1) -- (1,0.1) -- (0.6, 0.1) -- (0, 0.7) -- (0,1) -- (0.1, 1);
        \draw [blue, very thick] (0, 1) -- (0.9, 0.1);

        \draw [black, fill=green] (0, 1.02) rectangle (1, 1.2);
        \draw (-0.2, 0.5) node {$A - \textcolor{blue}{\sigma_{\ell_1}}I$};
        \draw (-0.2, 1.1) node {\textcolor{blue}{$C$}};

        \draw [black, fill=blue] (0.6, 1.02) rectangle (1, 1.2);
        \draw [black, fill=blue] (0.6, 0.1) rectangle (1, 1);

        \draw [fill=blue] (1, 0.1) -- (0.9, 0.1) -- (1, 0);

        \draw [fill=black, fill opacity=0.3] (0.6, 0) rectangle (1, 1.2);
        \draw [black] (0.6, 0.1) -- (1, 0.1);

        \draw (1, 0.05) node[anchor=west] {$\textcolor{blue}{R_{block}^{(\ell_1)}}$};

        \draw [fill=blue] (1.05, 0.3) rectangle (1.25, 0.5);
        \draw (1.01, 0.58) node[anchor=west] {$\textcolor{blue}{P^{(\ell_1)^*}}$};

        \draw (0.8, 1.3) node {$\textcolor{blue}{Z^{(\ell_1)}}$};
\begin{scope}[shift={(2.5,0)}]
        \draw [black, fill=green] (0.1, 1) -- (1, 1) -- (1,0.1) -- (0.6, 0.1) -- (0, 0.7) -- (0,1) -- (0.1, 1);
        \draw [red, very thick] (0, 1) -- (0.9, 0.1);

        \draw [black, fill=green] (0, 1.02) rectangle (1, 1.2);
        \draw (-0.2, 0.5) node {$A - \textcolor{red}{\sigma_{\ell_2}}I$};
        \draw (-0.2, 1.1) node {\textcolor{red}{$C$}};

        \draw [black, fill=red] (0.6, 1.02) rectangle (1, 1.2);
        \draw [black, fill=red] (0.6, 0.1) rectangle (1, 1);

        \draw [fill=red] (1, 0.1) -- (0.9, 0.1) -- (1, 0);

        \draw [fill=black, fill opacity=0.3] (0.6, 0) rectangle (1, 1.2);
        \draw [black] (0.6, 0.1) -- (1, 0.1);

        \draw (1, 0.05) node[anchor=west] {$\textcolor{red}{R_{block}^{(\ell_2)}}$};

        \draw [fill=red] (1.05, 0.3) rectangle (1.25, 0.5);
        \draw (1.01, 0.58) node[anchor=west] {$\textcolor{red}{P^{(\ell_2)^*}}$};

        \draw (0.8, 1.3) node {$\textcolor{red}{Z^{(\ell_2)}}$};
\end{scope}
\end{tikzpicture}

\noindent
\emph{Step 3:} Subsequently, the sliding window $Z^{(\ell)}$ moves by $n_b$ columns to the left.

\begin{tikzpicture}[xscale=3, yscale=3]
        \draw [black, fill=green] (0.1, 1) -- (1, 1) -- (1,0.1) -- (0.6, 0.1) -- (0, 0.7) -- (0,1) -- (0.1, 1);
        \draw [blue, very thick] (0, 1) -- (0.9, 0.1);

        \draw [black, fill=green] (0, 1.02) rectangle (1, 1.2);
        \draw (-0.2, 0.5) node {$A - \textcolor{blue}{\sigma_{\ell_1}}I$};
        \draw (-0.2, 1.1) node {\textcolor{blue}{$C$}};

        \draw [black, fill=blue] (0.6, 1.02) rectangle (1, 1.2);
        \draw [black, fill=blue] (0.6, 0.1) rectangle (1, 1);

        \draw [black, fill=blue] (1, 0.1) -- (0.9, 0.1) -- (1, 0) -- (1, 0.1); 

        \draw [fill=black, fill opacity=0.3] (0.5, 0.1) rectangle (0.9, 1.2);
        \draw [black] (0.5, 0.2) -- (0.9, 0.2);

        \draw (1, 0.05) node[anchor=west] {$\textcolor{blue}{R_{block}^{(\ell_1)}}$};

        \draw [fill=blue] (1.05, 0.3) rectangle (1.25, 0.5);
        \draw (1.01, 0.58) node[anchor=west] {$\textcolor{blue}{P^{(\ell_1)^*}}$};

        \draw (0.7, 1.3) node {$\textcolor{blue}{Z^{(\ell_1)}}$};
        \draw [blue, thick] (0.9, 0.1) -- (1, 0.1); 
\begin{scope}[shift={(2.5,0)}]
        \draw [black, fill=green] (0.1, 1) -- (1, 1) -- (1,0.1) -- (0.6, 0.1) -- (0, 0.7) -- (0,1) -- (0.1, 1);
        \draw [red, very thick] (0, 1) -- (0.9, 0.1);

        \draw [black, fill=green] (0, 1.02) rectangle (1, 1.2);
        \draw (-0.2, 0.5) node {$A - \textcolor{red}{\sigma_{\ell_2}}I$};
        \draw (-0.2, 1.1) node {\textcolor{red}{$C$}};

        \draw [black, fill=red] (0.6, 1.02) rectangle (1, 1.2);
        \draw [black, fill=red] (0.6, 0.1) rectangle (1, 1);

        \draw [black, fill=red] (1, 0.1) -- (0.9, 0.1) -- (1, 0) -- (1, 0.1); 

        \draw [fill=black, fill opacity=0.3] (0.5, 0.1) rectangle (0.9, 1.2);
        \draw [black] (0.5, 0.2) -- (0.9, 0.2);

        \draw (1, 0.05) node[anchor=west] {$\textcolor{red}{R_{block}^{(\ell_2)}}$};

        \draw [fill=red] (1.05, 0.3) rectangle (1.25, 0.5);
        \draw (1.01, 0.58) node[anchor=west] {$\textcolor{red}{P^{(\ell_2)^*}}$};

        \draw (0.7, 1.3) node {$\textcolor{red}{Z^{(\ell_2)}}$};
        \draw [red, thick] (0.9, 0.1) -- (1, 0.1); 
\end{scope}
\end{tikzpicture}

\noindent
Since we are only interested in obtaining the top-left $m\times m$ corner of $R_\ell$, in the rest of the algorithm we do not need the computed RQ factors $R_{block}^{(\ell)}$ and $P^{(\ell)}$ any more.
The same is true for the $n_{b}$ columns that dropped out of the sliding window $Z^{(\ell)}$ after it was moved. Therefore, when updating $Z^{(\ell)}$ in Step 2, only its first $m$ columns actually need to be updated, and, consequently, only the first $m$ columns of $P^{(\ell)^*}$ are needed to execute the update.

\begin{tikzpicture}[xscale=3, yscale=3]
        \draw [black, fill=green] (0.1, 1) -- (0.9, 1) -- (0.9, 0.1) -- (0.6, 0.1) -- (0, 0.7) -- (0,1) -- (0.1, 1);
        \draw [blue, very thick] (0, 1) -- (0.9, 0.1);

        \draw [black, fill=green] (0, 1.02) rectangle (0.9, 1.2);
        \draw (-0.2, 0.5) node {$A - \textcolor{blue}{\sigma_{\ell_1}}I$};
        \draw (-0.2, 1.1) node {\textcolor{blue}{$C$}};

        \draw [black, fill=blue] (0.6, 1.02) rectangle (0.9, 1.2);
        \draw [black, fill=blue] (0.6, 0.1) rectangle (0.9, 1);

        \draw [black, fill=white] (0.9, 1.02) rectangle (1, 1.2);
        \draw [black, fill=white] (0.9, 0.1) rectangle (1, 1);

        \draw [black, fill=white] (1, 0.1) -- (0.9, 0.1) -- (1, 0) -- (1, 0.1); 

        \draw [fill=black, fill opacity=0.3] (0.5, 0.1) rectangle (0.9, 1.2);
        \draw [black] (0.5, 0.2) -- (0.9, 0.2);

        \draw (1, 0.05) node[anchor=west] {$\textcolor{blue}{R_{block}^{(\ell_1)}}$};

        \draw [fill=white] (1.05, 0.3) rectangle (1.15, 0.5);
        \draw (1, 0.58) node[anchor=west] {$\textcolor{blue}{P^{(\ell_1)^*}}$};

        \draw (0.7, 1.3) node {$\textcolor{blue}{Z^{(\ell_1)}}$};
        \draw [white, thick] (0.9, 0.1) -- (1, 0.1); 
\begin{scope}[shift={(2.5,0)}]
        \draw [black, fill=green] (0.1, 1) -- (0.9, 1) -- (0.9, 0.1) -- (0.6, 0.1) -- (0, 0.7) -- (0,1) -- (0.1, 1);
        \draw [red, very thick] (0, 1) -- (0.9, 0.1);

        \draw [black, fill=green] (0, 1.02) rectangle (0.9, 1.2);
        \draw (-0.2, 0.5) node {$A - \textcolor{red}{\sigma_{\ell_2}}I$};
        \draw (-0.2, 1.1) node {\textcolor{red}{$C$}};

        \draw [black, fill=red] (0.6, 1.02) rectangle (0.9, 1.2);
        \draw [black, fill=red] (0.6, 0.1) rectangle (0.9, 1);

        \draw [black, fill=white] (0.9, 1.02) rectangle (1, 1.2);
        \draw [black, fill=white] (0.9, 0.1) rectangle (1, 1);

        \draw [black, fill=white] (1, 0.1) -- (0.9, 0.1) -- (1, 0) -- (1, 0.1); 

        \draw [fill=black, fill opacity=0.3] (0.5, 0.1) rectangle (0.9, 1.2);
        \draw [black] (0.5, 0.2) -- (0.9, 0.2);

        \draw (1, 0.05) node[anchor=west] {$\textcolor{red}{R_{block}^{(\ell_2)}}$};

        \draw [fill=white] (1.05, 0.3) rectangle (1.15, 0.5);
        \draw (1, 0.58) node[anchor=west] {$\textcolor{red}{P^{(\ell_2)^*}}$};

        \draw (0.7, 1.3) node {$\textcolor{red}{Z^{(\ell_2)}}$};
        \draw [white, thick] (0.9, 0.1) -- (1, 0.1); 
\end{scope}
\end{tikzpicture}

\newcommand{\Full}[3]{%
	\begin{tikzpicture}{0}{0}{#3}{#2}
		\draw [black, fill=#1] (0,0) rectangle (#3,#2);
	\end{tikzpicture}}

\newcommand{\Patt}[4]{%
	\begin{tikzpicture}[xscale=#4, yscale=#3]
		\draw [black, fill=#1] (0, 0) rectangle (1, 1);
		\draw [pattern=#2] (0, 0) rectangle (1, 1);
	\end{tikzpicture}}

\noindent
We now iterate Steps 1--3 in a loop. However, by observing one property of the moving sliding window, we can perform Step 2 far more efficiently. Note that in Step 3, the first $n_{b}$ columns of $Z^{(\ell)}$ are refilled with the original elements of $A-\sigma_\ell I$ and $C$.
These elements are same for all shifts, and they will be accessed only once. Therefore, $Z^{(\ell )}$ may be split into two parts:
\begin{enumerate}
    \item
        The first part consists of the first $n_{b}$ columns of $Z^{(\ell )}$ and is denoted by $Z_{1}$.
        This part is (almost) the same for all shifts and relates to the original elements of $A$ and $C$.
        The ``small'' differences relate to diagonal elements, and will be dealt with separately.
    \item
        The second part consists of the last $m$ columns of $Z^{(\ell )}$ that are specific to the shift $\sigma_{\ell}$,
        and is denoted by $Z_{2}^{(\ell )}$. These submatrices are the result of the update from Step 2,
        and are different for different shifts.
\end{enumerate}

\noindent
We can therefore reformulate Step 2 in the following way: \\

\noindent
\emph{Step 2, revisited:}
The update of the first $m$ columns of $Z_{1}$ for every shift is split into two parts and the result is stored in $Z_{2}^{(\ell )}$ for the next step:
\begin{enumerate}
    \item
        Batched GEMM is applied to obtain $Z_{2}^{(\ell )}=Z_{2}^{(\ell )}\cdot P^{(\ell)^*}(n_{b}+1:n_{b}+m,1:m)$:
        $$
            \textcolor{blue}{Z_2^{(\ell_1)}} = \Full{blue}{.5}{.2} \cdot \Patt{blue}{north west lines}{.2}{.2},\quad
            \textcolor{red}{Z_2^{(\ell_2)}} = \Full{red}{.5}{.2} \cdot \Patt{red}{north west lines}{.2}{.2}.
        $$
    \item
        A single additional GEMM is applied to obtain
        \begin{align*}
            \left[ \; Z_{2}^{(1)} \; \right. & \left. \; Z_{2}^{(2)} \quad \cdots \quad Z_{2}^{(s)} \; \right]
                = \left[\; Z_{2}^{(1)} \quad Z_{2}^{(2)} \quad \cdots \quad Z_{2}^{(s)} \; \right] \\
                    &+ Z_{1}\cdot \left[ \; P^{(1)^*}(1:n_{b},1:m) \quad P^{(2)^*}(1:n_{b},1:m) \quad \cdots \quad P^{(s)^*}(1:n_{b},1:m) \; \right]:
        \end{align*}
        $$
            \left[\begin{smallmatrix} \textcolor{blue}{Z_{2}^{(\ell_1)}} & \textcolor{red}{Z_{2}^{(\ell_2)}} \end{smallmatrix}\right]
                = \left[\begin{smallmatrix} \textcolor{blue}{Z_{2}^{(\ell_1)}} & \textcolor{red}{Z_{2}^{(\ell_2)}} \end{smallmatrix}\right]
                + \Full{green}{.5}{.2} \cdot
                    \left[\begin{smallmatrix} \Patt{blue}{north east lines}{.2}{.2} & \Patt{red}{north east lines}{.2}{.2} \end{smallmatrix}\right].
        $$
\end{enumerate}

\begin{tikzpicture}[xscale=3, yscale=3]
        \draw [black, fill=green] (0.1, 1) -- (0.8, 1) -- (0.8, 0.2) -- (0.5, 0.2) -- (0, 0.7) -- (0, 1) -- (0.1, 1);
        \draw [blue, very thick] (0, 1) -- (0.8, 0.2);

        \draw [black, fill=green] (0, 1.02) rectangle (0.9, 1.2);
        \draw (-0.2, 0.5) node {$A - \textcolor{blue}{\sigma_{\ell_1}}I$};
        \draw (-0.2, 1.1) node {\textcolor{blue}{$C$}};

        \draw [black, fill=blue] (0.6, 1.02) rectangle (0.9, 1.2);
        \draw [black, fill=blue] (0.6, 0.2) rectangle (0.9, 1);

        \draw [black, fill=white] (0.9, 1.02) rectangle (1, 1.2);
        \draw [black, fill=white] (0.9, 0.1) rectangle (1, 1);

        \draw [black, fill=white] (1, 0.1) -- (0.9, 0.1) -- (1, 0) -- (1, 0.1); 
        \draw [black, fill=blue] (0.9, 0.2) -- (0.8, 0.2) -- (0.9, 0.1) -- (0.9, 0.2); 

        \draw [fill=black, fill opacity=0.3] (0.5, 0.1) rectangle (0.9, 1.2);
        \draw [black] (0.5, 0.2) -- (0.9, 0.2);

        \draw (1, 0.05) node[anchor=west] {$\textcolor{blue}{R_{block}^{(\ell_1)}}$};

        \draw [fill=blue] (1.05, 0.3) rectangle (1.15, 0.5); 
        \draw [pattern=north west lines] (1.05, 0.3) rectangle (1.15, 0.4);
        \draw [pattern=north east lines] (1.05, 0.4) rectangle (1.15, 0.5);
        \draw (1, 0.58) node[anchor=west] {$\textcolor{blue}{P^{(\ell_1)^*}}$};

        \draw (0.7, 1.3) node {$\textcolor{blue}{Z^{(\ell_1)}}$};
        \draw [white, thick] (0.9, 0.1) -- (1, 0.1); 
\begin{scope}[shift={(2.5,0)}]
        \draw [black, fill=green] (0.1, 1) -- (0.8, 1) -- (0.8, 0.2) -- (0.5, 0.2) -- (0, 0.7) -- (0, 1) -- (0.1, 1);
        \draw [red, very thick] (0, 1) -- (0.8, 0.2);

        \draw [black, fill=green] (0, 1.02) rectangle (0.9, 1.2);
        \draw (-0.2, 0.5) node {$A - \textcolor{red}{\sigma_{\ell_2}}I$};
        \draw (-0.2, 1.1) node {\textcolor{red}{$C$}};

        \draw [black, fill=red] (0.6, 1.02) rectangle (0.9, 1.2);
        \draw [black, fill=red] (0.6, 0.2) rectangle (0.9, 1);

        \draw [black, fill=white] (0.9, 1.02) rectangle (1, 1.2);
        \draw [black, fill=white] (0.9, 0.1) rectangle (1, 1);

        \draw [black, fill=white] (1, 0.1) -- (0.9, 0.1) -- (1, 0) -- (1, 0.1); 
        \draw [black, fill=red] (0.9, 0.2) -- (0.8, 0.2) -- (0.9, 0.1) -- (0.9, 0.2); 

        \draw [fill=black, fill opacity=0.3] (0.5, 0.1) rectangle (0.9, 1.2);
        \draw [black] (0.5, 0.2) -- (0.9, 0.2);

        \draw (1, 0.05) node[anchor=west] {$\textcolor{red}{R_{block}^{(\ell_2)}}$};

        \draw [fill=red] (1.05, 0.3) rectangle (1.15, 0.5); 
        \draw [pattern=north west lines] (1.05, 0.3) rectangle (1.15, 0.4);
        \draw [pattern=north east lines] (1.05, 0.4) rectangle (1.15, 0.5);
        \draw (1, 0.58) node[anchor=west] {$\textcolor{red}{P^{(\ell_2)^*}}$};

        \draw (0.7, 1.3) node {$\textcolor{red}{Z^{(\ell_2)}}$};
        \draw [white, thick] (0.9, 0.1) -- (1, 0.1); 
\end{scope}
\end{tikzpicture}

Eventually, the sliding window will move as far as to include the first column of the matrix $\left[ \begin{smallmatrix} C\\ A-\sigma_\ell I \end{smallmatrix}\right]$.
For this last block,
the first $m$ columns of the matrix $\left[ \begin{smallmatrix} C\\ A-\sigma_\ell I \end{smallmatrix}\right]$ are
processed separately in non-blocked fashion.
The operations are structured in a way that each step of the reduction is followed by one step of the backward substitutions
for solving the system $R_\ell (1:m,1:m)^{-1} B(1:m,1:m)$.

\subsection{A GPU algorithm for transfer function evaluation}
\begin{algorithm}[h!]
\KwIn{$(A,B,C)\in\mathbb{R}^{n\times n} \times \mathbb{R}^{n\times m}\times \mathbb{R}^{p\times n}$
($(A,B)$ in the controller Hessenberg form); $\sigma\in\mathbb{C}^{s}$, and block dimension $n_{b}$}
\KwOut{$G\in \mathbb{C}^{p\times s\cdot m}$, where $G(1:p,(\ell -1)m+1:\ell m)=\mathcal{G}(\sigma(\ell ))$, $\ell = 1,\ldots ,s$}

\vspace*{0.3cm}
Allocate the following arrays on the GPU: \\
\tab $\mathbf{dZ_{1}}\in \mathbb{C}^{(p+n)\times n_b}$; denote $Z_{1} \equiv \mathbf{dZ_{1}}$\;
\tab $\mathbf{dZ_{2}}\in \mathbb{C}^{(p+n)\times s\cdot m}$; denote $Z_{2}^{(\ell)} \equiv \mathbf{dZ_{2}}(:,(\ell -1)m+1:\ell m)$\;
\tab $\mathbf{dZ_{block}}\in \mathbb{C}^{n_b\times s\cdot (m+n_b)}$; denote $Z_{block}^{(\ell)} \equiv \mathbf{dZ_{block}}(:,(\ell -1)(m+n_b)+1:\ell (m+n_b))$\;
\tab $\mathbf{dP}\in \mathbb{C}^{(n_{b}+m)\times s\cdot m}$; denote $P^{(\ell )^*}(1:n_{b}+m,1:m) \equiv \mathbf{dP}(:,(\ell -1)m+1:\ell m)$\;
\tab $\mathbf{dW}\in \mathbb{C}^{(p+n)\times s\cdot m}$\;

\vspace*{0.3cm}

\tcp{The following is executed on the GPU.}
\For{$\ell =1,2,\ldots ,s$}
{
    Copy the last $m$ columns of $\left[ \begin{array}{c} C\\ A \end{array}\right]$ to $Z_{2}^{(\ell )}$\;
}
Run a GPU kernel to subtract $\sigma_\ell$ from the diagonal of $Z_{2}^{(\ell )}$, for all $\ell=1, \ldots, s$\;
\For{$k=n,n-n_{b},n-2n_{b},\ldots m+1$}
{
    $n_{b}=\min \{ n_{b}, k-m\}$; $mn_{b}=\min \{ m, n_{b}\}$; $r=p+k-n_{b}$\;
    Copy $\left[ \begin{array}{c} C\\ A \end{array}\right](1:p+k,k-m-n_{b}+1:k-m)$ to $Z_{1}(1:p+k,1:n_{b})$\;
    \For{$\ell=1,2,\ldots ,s$}
    {
        Copy $Z_{1}(r+1:p+k,1:n_{b})$ to $Z_{block}^{(\ell )}(1:n_{b},1:n_{b})$\;
        Copy $Z_{2}^{(\ell )}(r+1:p+k,1:m)$ to $Z_{block}^{(\ell )}(1:n_{b},n_{b}+1:n_{b}+m)$\;
    }
    \If{$n_{b}>m$}
    {
        Run a GPU kernel to subtract $\sigma_\ell$ from the remaining diagonal elements in $Z_{block}^{(\ell)}$, for all $\ell=1, \ldots, s$\;
    }
    Run a GPU kernel to compute the RQ factorization for $Z_{block}^{(\ell )}$, producing $P^{(\ell )^*}(1:n_{b}+m,1:m)$, for all $\ell=1\ldots s$\;
    Perform batched multiplication: for all $\ell =1,\ldots, s$ in parallel compute
    	$\mathbf{dW}(1:r,(\ell -1)m+1:\ell m)=Z_{2}^{(\ell )}(1:r,1:m)\cdot P^{(\ell )^*}(n_{b}+1:n_{b}+m,1:m)$\;
    Copy $\mathbf{dW}(1:r,1:s\cdot m)$ to $\mathbf{{d}Z_{2}}(1:r,1:s\cdot m)$\;
    Run a GPU kernel to update $Z_{2}^{(\ell )}$, for all $\ell=1, \ldots, s$:
    	$Z_{2}^{(\ell )}(r-m+1:r-m+mn_{b},1:m)=Z_{2}^{(\ell )}(r-m+1:r-m+mn_{b},1:m)-$ $-\sigma (\ell )P^{(\ell )^*}(1:mn_{b},1:m)$\;
    $\mathbf{{d}Z_{2}}(1:r,1:s\cdot m) = \mathbf{{d}Z_{2}}(1:r,1:s\cdot m)+\mathbf{{d}Z_{1}}(1:r,1:n_{b})\cdot \mathbf{{d}P}(1:n_{b},1:s\cdot m) $\;
}
Copy $\mathbf{dZ_{2}}(1:p+m,1:s\cdot m)$ from the GPU to $\mathbf{Z_{2}}(1:p+m,1:s\cdot m)$ on the CPU\;
Revert the notation: $Z_{2}^{(\ell)} \equiv \mathbf{Z_{2}}(:,(\ell -1)m+1:\ell m)$\;

\vspace*{0.3cm}

\tcp{The following is executed on the CPU.}
\For{$\ell =1,2,\ldots s$}
{
    Reduce $Z_{2}^{(\ell )}(p+1:p+m,1:m)$ to the triangular form $\hat{T}(\ell )$ and simultaneously solve $\hat{T}(\ell )\hat{X}(\ell )=\hat{B}$, on CPU\;
    $G(1:p,(\ell -1)m+1:\ell m)= -Z_{2}^{(\ell )}(1:p,1:m)\cdot \hat{X}(\ell )$\;
}

\caption{Parallel GPU/CPU algorithm for computing $\mathcal{G}(\sigma (\ell ))=C(\sigma (\ell ) I - A)^{-1}B$, $\ell =1,\ldots,s$.}\label{alg-parallel-main}
\end{algorithm}


The ideas presented in the previous section are put together as Algorithm \ref{alg-parallel-main}. The vast majority of the algorithm is executed on the GPU, leaving only the final solution of the small-dimensional triangular systems to the CPU. We now explain some of the more important details of the algorithm.

The batched multiplication in line 19 may be computed by a single call to \alg{cublasZgemmBatched}, or by running multiple cuBLAS streams, each of which does one call to \alg{cublasZgemm}.

Algorithm \ref{alg-parallel-main} requires implementation of several GPU kernels (lines 9, 17, 18, 21). Each of these kernels has an obvious parallelization scheme: the tasks that have to be computed are the same for all shifts, and at the same time they are completely independent. Therefore, each thread block of the kernel will compute the particular task for one shift. The parallelization within each block, i.e.~deciding on the job for each thread within a block is trivial, except for the kernel computing the RQ factorization, which we now discuss in more detail.

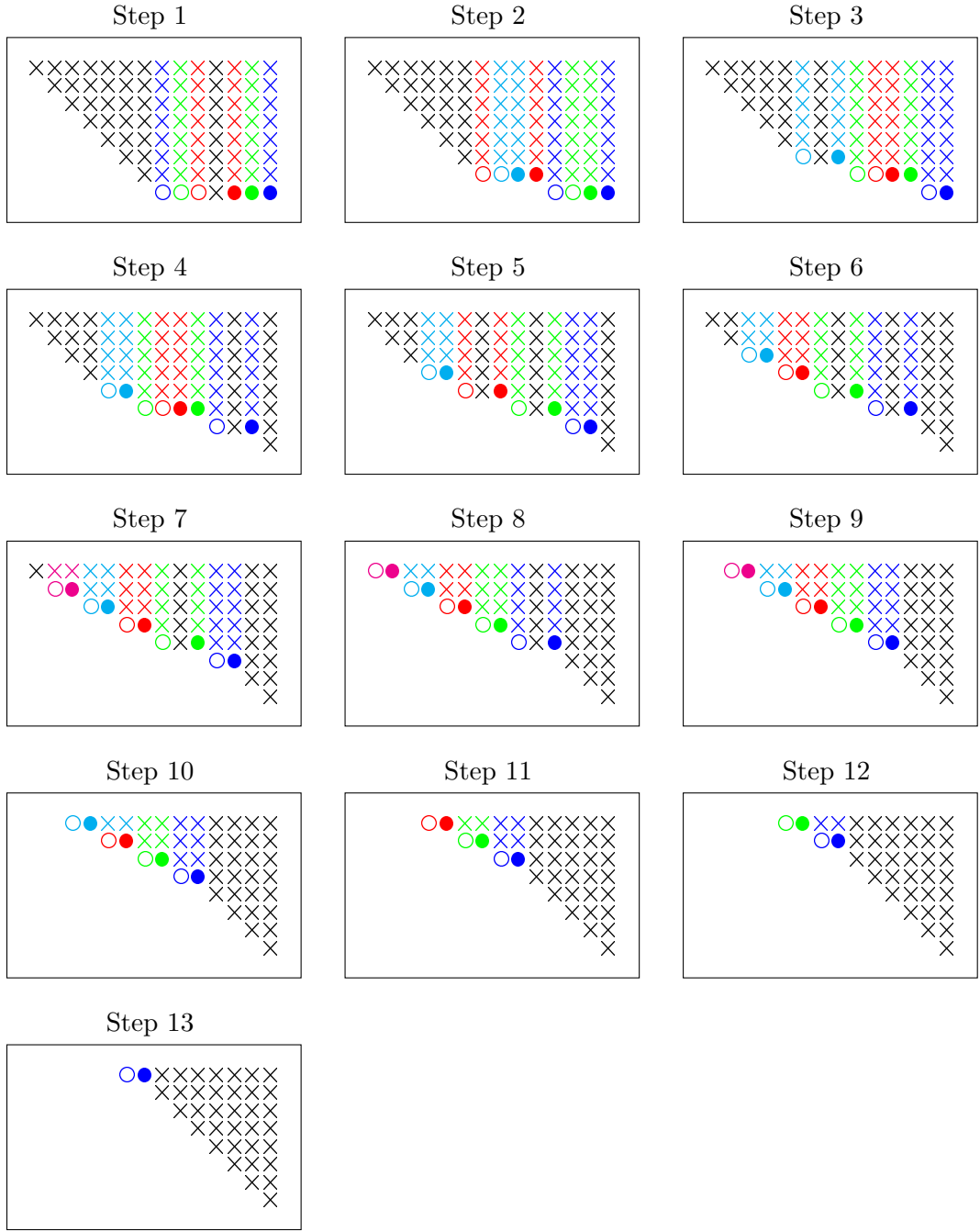
\begin{figure}[htbp]

\begin{center}
\begin{tikzpicture}[x=3.15em,y=3.15em]
\newcommand{\xx}{\node{\small \SmallCross};}
\newcommand{\bx}{\node[blue]{\small \SmallCross};}
\newcommand{\gx}{\node[green]{\small \SmallCross};}
\newcommand{\rx}{\node[red]{\small \SmallCross};}
\newcommand{\cx}{\node[cyan]{\small \SmallCross};}
\newcommand{\mx}{\node[magenta]{\small \SmallCross};}
\newcommand{\wx}{\node[white]{\small \SmallCross};}

\newcommand{\bo}{\node[blue]{\small \SmallCircle};}
\newcommand{\go}{\node[green]{\small \SmallCircle};}
\newcommand{\ro}{\node[red]{\small \SmallCircle};}
\newcommand{\co}{\node[cyan]{\small \SmallCircle};}
\newcommand{\mo}{\node[magenta]{\small \SmallCircle};}

\newcommand{\by}{\node[blue]{\small \FilledSmallCircle};}
\newcommand{\gy}{\node[green]{\small \FilledSmallCircle};}
\newcommand{\ry}{\node[red]{\small \FilledSmallCircle};}
\newcommand{\cy}{\node[cyan]{\small \FilledSmallCircle};}
\newcommand{\my}{\node[magenta]{\small \FilledSmallCircle};}

\begin{scope}[shift={(0,12)}]
\matrix [matrix of nodes, draw=black, column sep=-0.35cm, row sep=-0.35cm, matrix anchor=south west] at (0,0)
{
     \xx & \xx & \xx & \xx & \xx & \xx & \xx & \bx & \gx & \rx & \xx & \rx & \gx & \bx \\ 
         & \xx & \xx & \xx & \xx & \xx & \xx & \bx & \gx & \rx & \xx & \rx & \gx & \bx \\ 
         &     & \xx & \xx & \xx & \xx & \xx & \bx & \gx & \rx & \xx & \rx & \gx & \bx \\ 
         &     &     & \xx & \xx & \xx & \xx & \bx & \gx & \rx & \xx & \rx & \gx & \bx \\ 
         &     &     &     & \xx & \xx & \xx & \bx & \gx & \rx & \xx & \rx & \gx & \bx \\ 
         &     &     &     &     & \xx & \xx & \bx & \gx & \rx & \xx & \rx & \gx & \bx \\ 
         &     &     &     &     &     & \xx & \bx & \gx & \rx & \xx & \rx & \gx & \bx \\ 
         &     &     &     &     &     &     & \bo & \go & \ro & \xx & \ry & \gy & \by \\ 
};
\draw (1.7,2.2) node[anchor=south] {Step 1};
\end{scope}

\begin{scope}[shift={(4,12)}]
\matrix [matrix of nodes, draw=black, column sep=-0.35cm, row sep=-0.35cm, matrix anchor=south west] at (0,0)
{
     \xx & \xx & \xx & \xx & \xx & \xx & \rx & \cx & \cx & \rx & \bx & \gx & \gx & \bx \\ 
         & \xx & \xx & \xx & \xx & \xx & \rx & \cx & \cx & \rx & \bx & \gx & \gx & \bx \\ 
         &     & \xx & \xx & \xx & \xx & \rx & \cx & \cx & \rx & \bx & \gx & \gx & \bx \\ 
         &     &     & \xx & \xx & \xx & \rx & \cx & \cx & \rx & \bx & \gx & \gx & \bx \\ 
         &     &     &     & \xx & \xx & \rx & \cx & \cx & \rx & \bx & \gx & \gx & \bx \\ 
         &     &     &     &     & \xx & \rx & \cx & \cx & \rx & \bx & \gx & \gx & \bx \\ 
         &     &     &     &     &     & \ro & \co & \cy & \ry & \bx & \gx & \gx & \bx \\ 
         &     &     &     &     &     &     &     &     &     & \bo & \go & \gy & \by \\ 
};
\draw (1.7,2.2) node[anchor=south] {Step 2};
\end{scope}

\begin{scope}[shift={(8,12)}]
\matrix [matrix of nodes, draw=black, column sep=-0.35cm, row sep=-0.35cm, matrix anchor=south west] at (0,0)
{
     \xx & \xx & \xx & \xx & \xx & \cx & \xx & \cx & \gx & \rx & \rx & \gx & \bx & \bx \\ 
         & \xx & \xx & \xx & \xx & \cx & \xx & \cx & \gx & \rx & \rx & \gx & \bx & \bx \\ 
         &     & \xx & \xx & \xx & \cx & \xx & \cx & \gx & \rx & \rx & \gx & \bx & \bx \\ 
         &     &     & \xx & \xx & \cx & \xx & \cx & \gx & \rx & \rx & \gx & \bx & \bx \\ 
         &     &     &     & \xx & \cx & \xx & \cx & \gx & \rx & \rx & \gx & \bx & \bx \\ 
         &     &     &     &     & \co & \xx & \cy & \gx & \rx & \rx & \gx & \bx & \bx \\ 
         &     &     &     &     &     &     &     & \go & \ro & \ry & \gy & \bx & \bx \\ 
         &     &     &     &     &     &     &     &     &     &     &     & \bo & \by \\ 
};
\draw (1.7,2.2) node[anchor=south] {Step 3};
\end{scope}

\begin{scope}[shift={(0,9)}]
\matrix [matrix of nodes, draw=black, column sep=-0.35cm, row sep=-0.35cm, matrix anchor=south west] at (0,0)
{
     \xx & \xx & \xx & \xx & \cx & \cx & \gx & \rx & \rx & \gx & \bx & \xx & \bx & \xx \\ 
         & \xx & \xx & \xx & \cx & \cx & \gx & \rx & \rx & \gx & \bx & \xx & \bx & \xx \\ 
         &     & \xx & \xx & \cx & \cx & \gx & \rx & \rx & \gx & \bx & \xx & \bx & \xx \\ 
         &     &     & \xx & \cx & \cx & \gx & \rx & \rx & \gx & \bx & \xx & \bx & \xx \\ 
         &     &     &     & \co & \cy & \gx & \rx & \rx & \gx & \bx & \xx & \bx & \xx \\ 
         &     &     &     &     &     & \go & \ro & \ry & \gy & \bx & \xx & \bx & \xx \\ 
         &     &     &     &     &     &     &     &     &     & \bo & \xx & \by & \xx \\ 
         &     &     &     &     &     &     &     &     &     &     &     &     & \xx \\ 
};

\draw (1.7,2.2) node[anchor=south] {Step 4};
\end{scope}

\begin{scope}[shift={(4,9)}]
\matrix [matrix of nodes, draw=black, column sep=-0.35cm, row sep=-0.35cm, matrix anchor=south west] at (0,0)
{
     \xx & \xx & \xx & \cx & \cx & \rx & \xx & \rx & \gx & \xx & \gx & \bx & \bx & \xx \\ 
         & \xx & \xx & \cx & \cx & \rx & \xx & \rx & \gx & \xx & \gx & \bx & \bx & \xx \\ 
         &     & \xx & \cx & \cx & \rx & \xx & \rx & \gx & \xx & \gx & \bx & \bx & \xx \\ 
         &     &     & \co & \cy & \rx & \xx & \rx & \gx & \xx & \gx & \bx & \bx & \xx \\ 
         &     &     &     &     & \ro & \xx & \ry & \gx & \xx & \gx & \bx & \bx & \xx \\ 
         &     &     &     &     &     &     &     & \go & \xx & \gy & \bx & \bx & \xx \\ 
         &     &     &     &     &     &     &     &     &     &     & \bo & \by & \xx \\ 
         &     &     &     &     &     &     &     &     &     &     &     &     & \xx \\ 
};
\draw (1.7,2.2) node[anchor=south] {Step 5};
\end{scope}

\begin{scope}[shift={(8,9)}]
\matrix [matrix of nodes, draw=black, column sep=-0.35cm, row sep=-0.35cm, matrix anchor=south west] at (0,0)
{
     \xx & \xx & \cx & \cx & \rx & \rx & \gx & \xx & \gx & \bx & \xx & \bx & \xx & \xx \\ 
         & \xx & \cx & \cx & \rx & \rx & \gx & \xx & \gx & \bx & \xx & \bx & \xx & \xx \\ 
         &     & \co & \cy & \rx & \rx & \gx & \xx & \gx & \bx & \xx & \bx & \xx & \xx \\ 
         &     &     &     & \ro & \ry & \gx & \xx & \gx & \bx & \xx & \bx & \xx & \xx \\ 
         &     &     &     &     &     & \go & \xx & \gy & \bx & \xx & \bx & \xx & \xx \\ 
         &     &     &     &     &     &     &     &     & \bo & \xx & \by & \xx & \xx \\ 
         &     &     &     &     &     &     &     &     &     &     &     & \xx & \xx \\ 
         &     &     &     &     &     &     &     &     &     &     &     &     & \xx \\ 
};
\draw (1.7,2.2) node[anchor=south] {Step 6};
\end{scope}

\begin{scope}[shift={(0,6)}]
\matrix [matrix of nodes, draw=black, column sep=-0.35cm, row sep=-0.35cm, matrix anchor=south west] at (0,0)
{
     \xx & \mx & \mx & \cx & \cx & \rx & \rx & \gx & \xx & \gx & \bx & \bx & \xx & \xx \\ 
         & \mo & \my & \cx & \cx & \rx & \rx & \gx & \xx & \gx & \bx & \bx & \xx & \xx \\ 
         &     &     & \co & \cy & \rx & \rx & \gx & \xx & \gx & \bx & \bx & \xx & \xx \\ 
         &     &     &     &     & \ro & \ry & \gx & \xx & \gx & \bx & \bx & \xx & \xx \\ 
         &     &     &     &     &     &     & \go & \xx & \gy & \bx & \bx & \xx & \xx \\ 
         &     &     &     &     &     &     &     &     &     & \bo & \by & \xx & \xx \\ 
         &     &     &     &     &     &     &     &     &     &     &     & \xx & \xx \\ 
         &     &     &     &     &     &     &     &     &     &     &     &     & \xx \\ 
};
\draw (1.7,2.2) node[anchor=south] {Step 7};
\end{scope}

\begin{scope}[shift={(4,6)}]
\matrix [matrix of nodes, draw=black, column sep=-0.35cm, row sep=-0.35cm, matrix anchor=south west] at (0,0)
{
     \mo & \my & \cx & \cx & \rx & \rx & \gx & \gx & \bx & \xx & \bx & \xx & \xx & \xx \\ 
         &     & \co & \cy & \rx & \rx & \gx & \gx & \bx & \xx & \bx & \xx & \xx & \xx \\ 
         &     &     &     & \ro & \ry & \gx & \gx & \bx & \xx & \bx & \xx & \xx & \xx \\ 
         &     &     &     &     &     & \go & \gy & \bx & \xx & \bx & \xx & \xx & \xx \\ 
         &     &     &     &     &     &     &     & \bo & \xx & \by & \xx & \xx & \xx \\ 
         &     &     &     &     &     &     &     &     &     &     & \xx & \xx & \xx \\ 
         &     &     &     &     &     &     &     &     &     &     &     & \xx & \xx \\ 
         &     &     &     &     &     &     &     &     &     &     &     &     & \xx \\ 
};

\draw (1.7,2.2) node[anchor=south] {Step 8};
\end{scope}

\begin{scope}[shift={(8,6)}]
\matrix [matrix of nodes, draw=black, column sep=-0.35cm, row sep=-0.35cm, matrix anchor=south west] at (0,0)
{
     \wx & \mo & \my & \cx & \cx & \rx & \rx & \gx & \gx & \bx & \bx & \xx & \xx & \xx \\ 
         &     &     & \co & \cy & \rx & \rx & \gx & \gx & \bx & \bx & \xx & \xx & \xx \\ 
         &     &     &     &     & \ro & \ry & \gx & \gx & \bx & \bx & \xx & \xx & \xx \\ 
         &     &     &     &     &     &     & \go & \gy & \bx & \bx & \xx & \xx & \xx \\ 
         &     &     &     &     &     &     &     &     & \bo & \by & \xx & \xx & \xx \\ 
         &     &     &     &     &     &     &     &     &     &     & \xx & \xx & \xx \\ 
         &     &     &     &     &     &     &     &     &     &     &     & \xx & \xx \\ 
         &     &     &     &     &     &     &     &     &     &     &     &     & \xx \\ 
};
\draw (1.7,2.2) node[anchor=south] {Step 9};
\end{scope}

\begin{scope}[shift={(0,3)}]
\matrix [matrix of nodes, draw=black, column sep=-0.35cm, row sep=-0.35cm, matrix anchor=south west] at (0,0)
{
     \wx & \wx & \co & \cy & \cx & \cx & \gx & \gx & \bx & \bx & \xx & \xx & \xx & \xx \\ 
         &     &     &     & \ro & \ry & \gx & \gx & \bx & \bx & \xx & \xx & \xx & \xx \\ 
         &     &     &     &     &     & \go & \gy & \bx & \bx & \xx & \xx & \xx & \xx \\ 
         &     &     &     &     &     &     &     & \bo & \by & \xx & \xx & \xx & \xx \\ 
         &     &     &     &     &     &     &     &     &     & \xx & \xx & \xx & \xx \\ 
         &     &     &     &     &     &     &     &     &     &     & \xx & \xx & \xx \\ 
         &     &     &     &     &     &     &     &     &     &     &     & \xx & \xx \\ 
         &     &     &     &     &     &     &     &     &     &     &     &     & \xx \\ 
};
\draw (1.7,2.2) node[anchor=south] {Step 10};
\end{scope}

\begin{scope}[shift={(4,3)}]
\matrix [matrix of nodes, draw=black, column sep=-0.35cm, row sep=-0.35cm, matrix anchor=south west] at (0,0)
{
     \wx & \wx & \wx & \ro & \ry & \gx & \gx & \bx & \bx & \xx & \xx & \xx & \xx & \xx \\ 
         &     &     &     &     & \go & \gy & \bx & \bx & \xx & \xx & \xx & \xx & \xx \\ 
         &     &     &     &     &     &     & \bo & \by & \xx & \xx & \xx & \xx & \xx \\ 
         &     &     &     &     &     &     &     &     & \xx & \xx & \xx & \xx & \xx \\ 
         &     &     &     &     &     &     &     &     &     & \xx & \xx & \xx & \xx \\ 
         &     &     &     &     &     &     &     &     &     &     & \xx & \xx & \xx \\ 
         &     &     &     &     &     &     &     &     &     &     &     & \xx & \xx \\ 
         &     &     &     &     &     &     &     &     &     &     &     &     & \xx \\ 
};
\draw (1.7,2.2) node[anchor=south] {Step 11};
\end{scope}

\begin{scope}[shift={(8,3)}]
\matrix [matrix of nodes, draw=black, column sep=-0.35cm, row sep=-0.35cm, matrix anchor=south west] at (0,0)
{
     \wx & \wx & \wx & \wx & \go & \gy & \bx & \bx & \xx & \xx & \xx & \xx & \xx & \xx \\ 
         &     &     &     &     &     & \bo & \by & \xx & \xx & \xx & \xx & \xx & \xx \\ 
         &     &     &     &     &     &     &     & \xx & \xx & \xx & \xx & \xx & \xx \\ 
         &     &     &     &     &     &     &     &     & \xx & \xx & \xx & \xx & \xx \\ 
         &     &     &     &     &     &     &     &     &     & \xx & \xx & \xx & \xx \\ 
         &     &     &     &     &     &     &     &     &     &     & \xx & \xx & \xx \\ 
         &     &     &     &     &     &     &     &     &     &     &     & \xx & \xx \\ 
         &     &     &     &     &     &     &     &     &     &     &     &     & \xx \\ 
};
\draw (1.7,2.2) node[anchor=south] {Step 12};
\end{scope}

\begin{scope}[shift={(0,0)}]
\matrix [matrix of nodes, draw=black, column sep=-0.35cm, row sep=-0.35cm, matrix anchor=south west] at (0,0)
{
     \wx & \wx & \wx & \wx & \wx & \bo & \by & \xx & \xx & \xx & \xx & \xx & \xx & \xx \\ 
         &     &     &     &     &     &     & \xx & \xx & \xx & \xx & \xx & \xx & \xx \\ 
         &     &     &     &     &     &     &     & \xx & \xx & \xx & \xx & \xx & \xx \\ 
         &     &     &     &     &     &     &     &     & \xx & \xx & \xx & \xx & \xx \\ 
         &     &     &     &     &     &     &     &     &     & \xx & \xx & \xx & \xx \\ 
         &     &     &     &     &     &     &     &     &     &     & \xx & \xx & \xx \\ 
         &     &     &     &     &     &     &     &     &     &     &     & \xx & \xx \\ 
         &     &     &     &     &     &     &     &     &     &     &     &     & \xx \\ 
};
\draw (1.7,2.2) node[anchor=south] {Step 13};
\end{scope}

\end{tikzpicture}
\end{center}

\caption{Greedy parallel annihilation strategy for $m=6$ and $n_{b}=8$.}\label{fig_greedy_example}
\end{figure}

The kernel in line 18 has to compute the RQ factorization of several relatively small matrices of dimension $n_b \times (n_b+m)$. As mentioned above, each factorization is run by a different block of threads. On the other hand, it is more difficult to achieve high parallelism within each block. Here we opted for an algorithm based on Givens rotations, custom tailored for the trapezoidal pattern of non-zero elements in the matrices $Z_{block}^{(\ell )}$.

The RQ factorization is done in a number of steps. In each step, several Givens rotations are computed so that certain elements of $Z_{block}^{(\ell )}$ become zero (``annihilated''). To be run in parallel, these Givens rotations have to be independent, i.e.~they need to operate on disjoint columns of $Z_{block}^{(\ell )}$; we want to find as many such rotations as possible in each step. To this end, we need to devise an annihilation strategy: which elements of the matrix will be annihilated in which step? We suggest a strategy based on a greedy algorithm \cite{cosnard1983complexite,modi1984alternative}; Figure \ref{fig_greedy_example} shows the 13 steps needed to factorize a small trapezoidal matrix in such a strategy computed by Algorithm \ref{alg-grredy-parallel-strategy}. In the figure, $\times$ denotes a non-zero element, $\circ$ denotes an element which will become zero in a particular step, and $\bullet$ of the same color denotes the paired element which is used in the Givens rotation in order to introduce the zero. The non-zero elements above each pair have to be updated with the rotation at the end of each step.

Note that the size of $Z_{block}^{(\ell )}$ remains the same throughout Algorithm \ref{alg-parallel-main}. This allows us to precompute the annihilation strategy in advance (on the CPU), store it in the constant memory at the GPU, and reuse it at all times. Algorithm \ref{alg-grredy-parallel-strategy} shows the computation of the annihilation strategy. The sequence of matrix elements on which the Givens rotations are to be applied is stored in the array \emph{rotInfo}. Elements of \emph{rotInfo} come in triplets $(r, c1, c2)$: the zero at position $(r, c1)$ in the matrix $Z_{block}^{(\ell )}$ is to be introduced with the help of the paired matrix element at position $(r, c2)$. The number of rotations in each step is given in the array \emph{jobSize}. Algorithm \ref{alg-grredy-parallel-strategy} finds as many independent Givens rotations as possible in each step; it can easily be generalized to function with any pattern of non-zeros in the original matrix.

Armed with the precomputed annihilation strategy, Algorithm \ref{alg-parallel-RQ} finally computes the RQ factorization. This kernel is configured to run with $s$ thread blocks, each containing many (say, $128$) threads. For each step of the strategy, the thread block first computes the sines and the cosines for all Givens rotations involved in the step, and stores them in the shared memory belonging to the block. This is done in parallel, but not all threads take part---only as many as there are rotations in the step. Once this is done, two matrices need to be updated with the rotations: $Z_{block}^{(\ell )}$ and $P^{(\ell )^*}$. All threads of the block participate in these updates, and the elements that need to be 
updated are assigned to the threads in a round-robin manner. This assignment is done in lines 12--13 for $Z_{block}^{(\ell )}$, and in lines 21--22 for $P^{(\ell )^*}$. The former is slightly more complicated then the latter, since all the rows in $P^{(\ell )^*}$ are updated {by} all rotations, which is not true for $Z_{block}^{(\ell )}$. By assigning the updates in the described way, the amount of computation that each thread has to do is roughly the same. Figure \ref{fig_parallel_RQ_updates} shows a small example of how this is done.

\begin{algorithm}[h!]
\KwIn{$\text{\emph{nRows}}$ and $\text{\emph{nCols}}$---the dimensions of the trapezoidal matrix}
\KwOut{Number of parallel steps $numSteps$, total number of Givens rotations $\text{\emph{numRots}}$, array $\text{\emph{jobSize}} \in \mathbb{N}^{\text{\emph{numSteps}}}$ with number of rotations determined in every step, and array $\text{\emph{rotInfo}} \in \mathbb{N}^{3\cdot \text{\emph{numRots}}}$ with indices $\text{\emph{r}}$, $\text{\emph{c1}}$ and $\text{\emph{c2}}$ for every rotation.}

\vspace*{0.3cm}
$\delta = \text{\emph{nCols}} - \text{\emph{nRows}}$\;
\tcp{Initialization}
Label all indices $(r,c)$ with $r\le c\le r+\delta$, as ready to be used in the algorithm\;
Label all indices $(r,c)$ with $r=\text{\emph{nRows}}$ and $c=r$, as available for the first step\;
$\text{\emph{numRots}}=0$; $\text{\emph{done}}=\text{\emph{false}}$; $\text{\emph{numSteps}}=1$\;
\While{not $\text{\emph{done}}$}
{
    $\text{\emph{done}} = \text{\emph{true}}$; $\text{\emph{jobSize(numSteps)}}=0$\;
    \tcp{Cleanup from the previous step.}
    Change labels of busy indices to ready\;
    Change labels of indices just above the annihilated ones to available\;
    \tcp{Find the rotations for the current step.}
    \For{$r=\text{\emph{nRows}},\text{\emph{nRows}}-1,\ldots ,1$}
    {
        \For{$c1=r,r+1,\ldots ,r+\delta$}
        {
            \If{$(r,c1)$ is labeled ready and available}
            {
                \tcp{Element at (r, c1) will be annihilated.}
                \For{$c2=r+\delta,r+\delta-1,\ldots ,c1+1$}
                {
                    \If{$(r,c2)$ is labeled ready and available}
                    {
                        \tcp{Element at (r, c2) will help to annihilate (r, c1).}
                        Store $r$, $c1$, $c2$ in $\text{\emph{rotInfo}}(3\cdot\text{\emph{numRots}}+1:3\cdot\text{\emph{numRots}}+3)$\;
                        $\text{\emph{numRots}}=\text{\emph{numRots}}+1$\;
                        $\text{\emph{jobSize(numSteps)}}=\text{\emph{jobSize(numSteps)}}+1$\;
                        Label $(r,c1)$ as not ready, and $(r,c2)$ as busy\;
                        \If{$r>1$}
                        {
                            Label $(r-1,c1)$ as not available\;
                        }
                        $\text{\emph{done}}=\text{\emph{false}}$\;
					    break\;
                    }
                }
            }
        }
    }
    $\text{\emph{numSteps}}=\text{\emph{numSteps}}+1$\;
}
\tcp{The last step was empty}
$\text{\emph{numSteps}}=\text{\emph{numSteps}}-1$\;

\caption{Greedy parallel annihilation strategy.}\label{alg-grredy-parallel-strategy}
\end{algorithm}

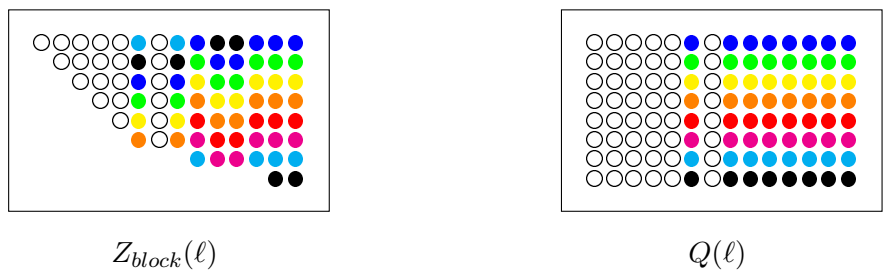
\begin{figure}[h!]

\begin{center}
\begin{tikzpicture}[x=3.15em,y=3.15em]
\newcommand{\xc}{\node{\small \SmallCircle};}
\newcommand{\xo}{\node{\small \FilledSmallCircle};}
\newcommand{\bo}{\node[blue]{\small \FilledSmallCircle};}
\newcommand{\go}{\node[green]{\small \FilledSmallCircle};}
\newcommand{\yo}{\node[yellow]{\small \FilledSmallCircle};}
\newcommand{\oo}{\node[orange]{\small \FilledSmallCircle};}
\newcommand{\ro}{\node[red]{\small \FilledSmallCircle};}
\newcommand{\co}{\node[cyan]{\small \FilledSmallCircle};}
\newcommand{\mo}{\node[magenta]{\small \FilledSmallCircle};}
\newcommand{\po}{\node[purple]{\small \FilledSmallCircle};}
\newcommand{\wo}{\node[white]{\small \FilledSmallCircle};}

\matrix [matrix of nodes, draw=black, column sep=-0.35cm, row sep=-0.35cm, matrix anchor=south west] at (0,0)
{
     \xc & \xc & \xc & \xc & \xc & \co & \xc & \co & \bo & \xo & \xo & \bo & \bo & \bo \\ 
         & \xc & \xc & \xc & \xc & \xo & \xc & \xo & \go & \bo & \bo & \go & \go & \go \\ 
         &     & \xc & \xc & \xc & \bo & \xc & \bo & \yo & \go & \go & \yo & \yo & \yo \\ 
         &     &     & \xc & \xc & \go & \xc & \go & \oo & \yo & \yo & \oo & \oo & \oo \\ 
         &     &     &     & \xc & \yo & \xc & \yo & \ro & \oo & \oo & \ro & \ro & \ro \\ 
         &     &     &     &     & \oo & \xc & \oo & \mo & \ro & \ro & \mo & \mo & \mo \\ 
         &     &     &     &     &     &     &     & \co & \mo & \mo & \co & \co & \co \\ 
         &     &     &     &     &     &     &     &     &     &     &     & \xo & \xo \\ 
};
\draw (1.7,-0.2) node[anchor=north] {$Z_{block}(\ell )$};

\begin{scope}[shift={(6,0)}]
\matrix [matrix of nodes, draw=black, column sep=-0.35cm, row sep=-0.35cm, matrix anchor=south west] at (0,0)
{
     \xc & \xc & \xc & \xc & \xc & \bo & \xc & \bo & \bo & \bo & \bo & \bo & \bo & \bo \\ 
     \xc & \xc & \xc & \xc & \xc & \go & \xc & \go & \go & \go & \go & \go & \go & \go \\ 
     \xc & \xc & \xc & \xc & \xc & \yo & \xc & \yo & \yo & \yo & \yo & \yo & \yo & \yo \\ 
     \xc & \xc & \xc & \xc & \xc & \oo & \xc & \oo & \oo & \oo & \oo & \oo & \oo & \oo \\ 
     \xc & \xc & \xc & \xc & \xc & \ro & \xc & \ro & \ro & \ro & \ro & \ro & \ro & \ro \\ 
     \xc & \xc & \xc & \xc & \xc & \mo & \xc & \mo & \mo & \mo & \mo & \mo & \mo & \mo \\ 
     \xc & \xc & \xc & \xc & \xc & \co & \xc & \co & \co & \co & \co & \co & \co & \co \\ 
     \xc & \xc & \xc & \xc & \xc & \xo & \xc & \xo & \xo & \xo & \xo & \xo & \xo & \xo \\ 
};
\draw (1.7,-0.2) node[anchor=north] {$Q(\ell )$};
\end{scope}

\end{tikzpicture}
\end{center}

\caption{Updating $Z_{block}^{(\ell)}$ and $P^{(\ell )}$ at the end of Step 3 from Figure \ref{fig_greedy_example}. We assume that the thread block contains $8$ threads. All elements of a particular color are updated by the same thread.}\label{fig_parallel_RQ_updates}
\end{figure}

\begin{algorithm}[h!]
\KwIn{$\text{\emph{nRows}}$, $\text{\emph{nCols}}$, $\text{\emph{numSteps}}$, $\text{\emph{jobSize}}$, $\text{\emph{rotInfo}}$, $\mathbf{{d}Z_{block}}\in \mathbb{C}^{\text{\emph{nRows}}\times (s\cdot \text{\emph{nCols}})}$ so that $Z_{block}^{(\ell)} \equiv\mathbf{{d}Z_{block}}(:,(\ell -1)\cdot\text{\emph{nCols}}+1:\ell \cdot \text{\emph{nCols}})$, $\mathbf{{d}P}\in \mathbb{C}^{\text{\emph{nCols}}\times (s\cdot \text{\emph{nCols}})}$ initialized to $s$ identity matrices, so that $P^{(\ell)^*} \equiv \mathbf{{d}P}(:,(\ell -1)\cdot \text{\emph{nCols}}+1:\ell \cdot \text{\emph{nCols}})$}
\KwOut{Orthogonal matrices from RQ factorizations stored in $\mathbf{{d}P}$}

\vspace*{0.3cm}
Prepare arrays $sh\_c$ and $sh\_s$ in shared memory for storing rotation cosines and sines\;
$\ell =\text{\emph{blockIdx}}.x$; $tid = \text{\emph{threadIdx}}.x$; $\text{\emph{offset}}=0$\;
\For{$\text{\emph{step}}=1,2,\ldots ,\text{\emph{numSteps}}-1$}
{
	\tcp{Compute the rotations' parameters.}
    \If{$\text{\emph{tid}} < \text{\emph{jobSize(step)}}$}
    {
        $\text{\emph{r}} = \text{\emph{rotInfo}}(\text{\emph{offset}} + 3\cdot \text{\emph{tid}}+1)$\;
        $\text{\emph{c1}} = \text{\emph{rotInfo}}(\text{\emph{offset}} + 3\cdot \text{\emph{tid}}+2)$;
        $\text{\emph{c2}} = \text{\emph{rotInfo}}(\text{\emph{offset}} + 3\cdot \text{\emph{tid}}+3)$\;
        Determine cosine $cc$ and sine $ss$ of the Givens rotation to annihilate $Z_{block}^{(\ell )}(\text{\emph{r}},\text{\emph{c1}})$ with the help of $Z_{block}^{(\ell )}(\text{\emph{r}},\text{\emph{c2}})$\;
        $sh\_c(\text{\emph{tid}})=cc$; $sh\_s(\text{\emph{tid}})=ss$\;
    }
    Synchronize threads\;
	\tcp{Update $Z_{block}^{(\ell)}$ with the computed rotations.}
    $rot = 0$; $r\_\text{\emph{begin}} = 0$\;
    \For{$r=\text{\emph{tid}},\text{\emph{tid}}+\text{\emph{blockDim}}.x,\ldots $}
    {
        \While{$\text{\emph{rot}} < \text{\emph{jobSize(step)}}$ \&\& $r \ge r\_\text{\emph{begin}} + \text{\emph{rotInfo}}(\text{\emph{offset}} + 3\cdot \text{\emph{rot}} +1)$}
        {
            $r\_\text{\emph{begin}} = r\_\text{\emph{begin}} + \text{\emph{rotInfo}}(\text{\emph{offset}} + 3\cdot \text{\emph{rot}} +1)$; $\text{\emph{rot}}=\text{\emph{rot}}+1$\;
        }
        \If{$\text{\emph{rot}} = \text{\emph{jobSize(step)}}$}
        {
				break\;
        }
        $r\_\text{\emph{local}} = r - r\_\text{\emph{begin}}$\;
        $\text{\emph{c1}} = \text{\emph{rotInfo}}(\text{\emph{offset}} + 3\cdot \text{\emph{rot}}+2)$; $\text{\emph{c2}} = \text{\emph{rotInfo}}(\text{\emph{offset}} + 3\cdot \text{\emph{rot}}+3)$\;
        Apply the Givens rotation $\text{\emph{rot}}$ determined by $cc = sh\_c(\text{\emph{rot}})$ and $ss = sh\_s(\text{\emph{rot}})$ to $Z_{block}^{(\ell)}(r\_\text{\emph{local}}, c1)$ and $Z_{block}^{(\ell)}(r\_\text{\emph{local}}, c2)$\;
    }
	\tcp{Update $P^{(\ell)}$ with the computed rotations.}
    $\text{\emph{rot}} = 0$; $r\_\text{\emph{begin}} = 0$\;
    \For{$r=\text{\emph{tid}},\text{\emph{tid}}+\text{\emph{blockDim}}.x,\ldots $}
    {
        \While{$\text{\emph{rot}} < \text{\emph{jobSize(step)}}$ \&\& $r \ge r\_\text{\emph{begin}} + \text{\emph{nCols}}$}
        {
            $r\_\text{\emph{begin}} = r\_\text{\emph{begin}} + \text{\emph{nCols}}$; $\text{\emph{rot}}=\text{\emph{rot}}+1$\;
        }
        \If{$\text{\emph{rot}} = \text{\emph{jobSize(step)}}$}
        {
				break\;
        }
        $r\_\text{\emph{local}} = r - r\_\text{\emph{begin}}$\;
        $\text{\emph{c1}} = \text{\emph{rotInfo}}(\text{\emph{offset}} + 3\cdot \text{\emph{rot}}+2)$; $\text{\emph{c2}} = \text{\emph{rotInfo}}(\text{\emph{offset}} + 3\cdot \text{\emph{rot}}+3)$\;
        Apply the Givens rotation $\text{\emph{rot}}$ determined by $cc = sh\_c(\text{\emph{rot}})$ and $ss = sh\_s(\text{\emph{rot}})$ to $P^{(\ell )^*}(r\_\text{\emph{local}}, c1)$ and $P^{(\ell )^*}(r\_\text{\emph{local}}, c2)$\;
    }
    $\text{\emph{offset}}=\text{\emph{offset}}+3\cdot \text{\emph{jobSize(step)}}$\;
    Synchronize threads\;
}

\caption{Parallel RQ factorizations of blocks $Z_{block}^{(\ell )}$.}\label{alg-parallel-RQ}
\end{algorithm}


\subsection{Algorithm for solving shifted systems with different right-hand sides for different shifts}\label{subs:irka:syst}
As mentioned in \S \ref{intro}, the core and most time consuming operation in the IRKA algorithm \cite{IRKA} is solving shifted systems, where in the case of multiple inputs and outputs the systems to be solved have different right-hand sides for different shifts.
These problems can be efficiently solved by a simple modification of Algorithm \ref{alg-parallel-main}.

Recall that the systems to be solved are of the following forms
\begin{align}
(A-\sigma_{\ell}I)x     &=\bb_{\ell},\label{solver:eq:origsyst}\\
(A-\sigma_{\ell}I)^{T}x &=\cc_{\ell}, \label{solver:eq:transsyst}
\end{align}
where the right-hand sides $\bb_{\ell}= B \widehat{\bb}_{\ell}$ and $\cc_{\ell}= C^T \widehat{\cc}_{\ell}$, $\ell=1,\ldots, s$, are in general complex.
%
As discussed in \S \ref{mhess}, we can assume that  $A$ and $B$ are already reduced to the controller Hessenberg form, so that  each right-hand side in the systems (\ref{solver:eq:origsyst}) has at most $m$ nontrivial elements. Hence, an  approach analogous to the one described in Algorithm \ref{alg-parallel-main} can be applied here as well. The main differences are at the end of the algorithm, when solving small $m\times m$ upper triangular systems, and are listed below.}
{
\begin{itemize}
\item The matrix $C$ is not required here and it is replaced by $Q_{\ell}^{*}$. At the beginning it is initialized with identity $I$, whose elements are just set to the values $0$ and $1$ in $Z^{(\ell )}$, not copied.
\item Therefore, the auxiliary array $Z^{(\ell )}$ will require $(2n)\times (n_{b}+m)$ elements for storing the current columns of the thus far transformed matrices $Q_{\ell}^{*}$ and $A-\sigma_\ell I$, organized as $\left[ \begin{smallmatrix} Q_{\ell}^{*}\\ A-\sigma_\ell I \end{smallmatrix}\right]$.
\item Line {26} solves systems with different right-hand sides for different shifts obtaining $y_{\ell}=R_{\ell}(1:m,1:m)^{-1}\bb_{\ell}(1:m)$.
\item Line {27} computes the final solution as $x_{\ell}=Q_{\ell}^{*}(1:n,1:m)y_{\ell}$.
\end{itemize}
}

{On the other hand, the transposed systems (\ref{solver:eq:transsyst}) have lower $m$-Hessenberg system matrices and unstructured right-hand sides. That means, that for every shift the system matrix is reduced to the lower triangular form by the LQ factorization $(A - \sigma_{\ell} I)^{T}=L_{\ell} Q_{\ell}$.
This reduction is performed in the similar way as the RQ factorization, except that the diagonal blocks are processed from top to bottom.}
{Since the right-hand sides have no particular form, computation of the solution requires all elements of matrices $L_{\ell}$ and $Q_{\ell}$.}

It turns out that it is still possible to retain most of the efficiency of Algorithm \ref{alg-parallel-main} with unreduced right-hand sides. The crucial role for
that is played by the {LQ} factorization.
Since the {LQ} factorization is reducing elements {top-down}, and since the {forward} substitution for
solving systems in {lower} triangular form is computing elements of the solution in the same fashion, all necessary updates of the final solution
are going to be performed on the fly, without ever forming complete factors $L_{\ell}$ and $Q_{\ell}$.
Therefore, the algorithm is still based on the three steps
\begin{enumerate}
\item solution of the $n\times n$ triangular system $y_{\ell} = { L_\ell^{-1} \cc_{\ell}}$,
\item computation of $Q_{\ell}^{*}$,
\item matrix multiplication $x_{\ell} = Q_{\ell}^{*} \cdot y_{\ell}$,
\end{enumerate}
but they are performed in parallel. As soon as an element of the solution $y_{\ell}$ of the triangular system  is obtained, and the appropriate column of $Q_{\ell}^{*}$ is computed, the final solution $x_{\ell}$ is updated with this data, and the data are then discarded. Thus, the basic operations
for  {$k=1,2,\ldots ,n$} are
\begin{enumerate}
\item obtain element {$L_{\ell}(k,k)$} by annihilating {$m$ superdiagonal elements in the $k$-th row of $(A-\sigma_{\ell}I)^{T}$},
\item compute element $y_{\ell}(k)$, {and update all other components of $y_{\ell}$ below,}
\item compute the $k$-th column of $Q_{\ell}^{*}$,
\item update $x_{\ell}=x_{\ell}+Q_{\ell}(1:n,k)\cdot y_{\ell}(k)$.
\end{enumerate}

The blocked version is organized in the similar way as  described in \S \ref{subsect-seq-alg-tfe}. Again, we will require the auxiliary {$(2n)\times (n_{b}+m)$} arrays $Z^{(\ell )}$ for storing the current columns of the thus far transformed matrices $Q_{\ell}^{*}$ and {$(A-\sigma_\ell I)^{T}$}, organized as {$\left[ \begin{smallmatrix} (A-\sigma_\ell I)^{T}\\Q_{\ell}^{*} \end{smallmatrix}\right]$}. $Z^{(\ell )}$ is further split into two parts, as before, {but $Z_{2}^{(\ell )}$ is now followed by $Z_{1}$}. Additionally we will need another auxiliary $(2n)\times 1$ arrays $w^{(\ell )}$ for storing current versions of {$y_{\ell}$ and $x_{\ell}$}, one above the other. It is possible to update these two vectors simultaneously with the same operations, by starting with $-I$ in the {lower} part of $Z^{(\ell)}$. The final computational procedure is presented in Algorithm \ref{alg-parallel-trans-diff-rhs}.

    \begin{tikzpicture}[xscale=2.5, yscale=2.5]
        \draw [black, fill=green] (0, 0) rectangle (1, 1);
        \draw [black, fill=green] (0, 1.02) -- (1, 1.02) -- (1,1.32) -- (0.4, 1.92) -- (0,1.92)  -- (0,1.02);
        \draw [blue, very thick] (0.1,1.92) -- (1, 1.02);

        \draw (-0.2, 0.5) node {\textcolor{blue}{$Q^{(\ell_{1})*}$}};
        \draw (-0.4, 1.51) node {$(A - \textcolor{blue}{\sigma_{\ell_{1}}}I)^{T}$};

        \draw [black, fill=blue] (0, 0) rectangle (0.4, 1);
        \draw [black, fill=blue] (0, 1.02) rectangle (0.4, 1.92);

        \draw [fill=blue] (0.1, 1.92) -- (0, 2.02) -- (0, 1.92);

        \draw [fill=black, fill opacity=0.3] (0, 0) rectangle (0.4, 2.02);
        \draw [black] (0, 1.92) -- (0.4, 1.92);

        \draw (-0.2, 1.97) node {$\textcolor{blue}{L_{block}^{(\ell_{1})}}$};

        \draw [fill=blue] (1.4, 1.72) rectangle (1.6, 1.92);
        \draw (1.5, 2.07) node {$\textcolor{blue}{P^{(\ell_{1})*}}$};

        \draw (0.2, 2.12) node {$\textcolor{blue}{Z^{(\ell_{1})}}$};

        \draw [black, fill=green] (1.05, 0) rectangle (1.1,1);
        \draw [black, fill=green] (1.05, 1.02) rectangle (1.1,1.92);
        \draw [fill=blue] (1.05, 1.92) rectangle (1.1,2.02);
        \draw [fill=black, fill opacity=0.3] (1.05, 0) rectangle (1.1,2.02);

        \draw (1.23, 0.5) node {$\textcolor{blue}{x_{\ell_{1}}}$};
        \draw (1.23, 1.52) node {$\textcolor{blue}{y_{\ell_{1}}}$};
        \draw (0.8, 1.97) node {$\textcolor{blue}{y_{\ell_{1},block}}$};

        \draw (1.07, 2.12) node {$\textcolor{blue}{w^{(\ell_{1})}}$};

    \begin{scope}[shift={(2.7,0)}]
        \draw [black, fill=green] (0, 0) rectangle (1, 1);
        \draw [black, fill=green] (0, 1.02) -- (1, 1.02) -- (1,1.32) -- (0.4, 1.92) -- (0,1.92)  -- (0,1.02);
        \draw [red, very thick] (0.1,1.92) -- (1, 1.02);

        \draw (-0.2, 0.5) node {\textcolor{red}{$Q^{(\ell_{2})*}$}};
        \draw (-0.4, 1.51) node {$(A - \textcolor{red}{\sigma_{\ell_{2}}}I)^{T}$};

        \draw [black, fill=red] (0, 0) rectangle (0.4, 1);
        \draw [black, fill=red] (0, 1.02) rectangle (0.4, 1.92);

        \draw [fill=red] (0.1, 1.92) -- (0, 2.02) -- (0, 1.92);

        \draw [fill=black, fill opacity=0.3] (0, 0) rectangle (0.4, 2.02);
        \draw [black] (0, 1.92) -- (0.4, 1.92);

        \draw (-0.2, 1.97) node {$\textcolor{red}{L_{block}^{(\ell_{2})}}$};

        \draw [fill=red] (1.4, 1.72) rectangle (1.6, 1.92);
        \draw (1.5, 2.07) node {$\textcolor{red}{P^{(\ell_{2})*}}$};

        \draw (0.2, 2.12) node {$\textcolor{red}{Z^{(\ell_{2})}}$};

        \draw [black, fill=green] (1.05, 0) rectangle (1.1,1);
        \draw [black, fill=green] (1.05, 1.02) rectangle (1.1,1.92);
        \draw [fill=red] (1.05, 1.92) rectangle (1.1,2.02);
        \draw [fill=black, fill opacity=0.3] (1.05, 0) rectangle (1.1,2.02);

        \draw (1.23, 0.5) node {$\textcolor{red}{x_{\ell_{2}}}$};
        \draw (1.23, 1.52) node {$\textcolor{red}{y_{\ell_{2}}}$};
        \draw (0.8, 1.97) node {$\textcolor{red}{y_{\ell_{2},block}}$};

        \draw (1.07, 2.12) node {$\textcolor{red}{w^{(\ell_{2})}}$};
    \end{scope}
    \end{tikzpicture}

\begin{algorithm}[h!]
\KwIn{$A\in\mathbb{R}^{n\times n}$ in {$m$-}Hessenberg form, {$\mathbf{c}\in \mathbb{C}^{n\times s}$} so that {$\mathbf{c}(:,\ell)=\cc_{\ell}$}, $\sigma\in\mathbb{C}^{s}$, and block dimension $n_{b}$}
\KwOut{$\mathbf{x}\in \mathbb{R}^{n\times s}$ so that $\mathbf{x}(:,\ell)=x_{\ell}$}

\vspace*{0.3cm}
{
Allocate the following arrays on the GPU: \\
\tab $\mathbf{dZ_{1}}\in \mathbb{C}^{2n\times n_b}$; denote $Z_{1} \equiv \mathbf{dZ_{1}}$\;
\tab $\mathbf{dZ_{2}}\in \mathbb{C}^{2n\times s\cdot m}$; denote $Z_{2}^{(\ell)} \equiv \mathbf{dZ_{2}}(:,(\ell -1)m+1:\ell m)$\;
\tab $\mathbf{dw}\in \mathbb{C}^{2n\times s}$; denote $w^{(\ell)} \equiv \mathbf{dw}(:,\ell)$\;
\tab $\mathbf{dZ_{block}}\in \mathbb{C}^{n_b\times s\cdot (m+n_b)}$; denote $Z_{block}^{(\ell)} \equiv \mathbf{dZ_{block}}(:,(\ell -1)(m+n_b)+1:\ell (m+n_b))$\;
\tab $\mathbf{dP}\in \mathbb{C}^{(n_{b}+m)\times s\cdot m}$; denote $P^{(\ell )^*}(1:n_{b}+m,n_{b}+1:n_{b}+m) \equiv \mathbf{dP}(:,(\ell -1)m+1:\ell m)$\;
\tab $\mathbf{dW}\in \mathbb{C}^{(p+n)\times s}$\;


\vspace*{0.3cm}

\tcp{The following is executed on the GPU.}
}
\For{$\ell =1,2,\ldots ,s$}
{
    Set {$w^{(\ell )}(1:n)=\cc(:,\ell)$} and {$w^{(\ell )}(n+1:2n)=0$}\;
    Fill $Z_{2}^{(\ell )}$ with the {first $m$ columns of $\left[ \begin{array}{c} A^{T}\\ -I \end{array}\right]$}\;
}
{
Run a GPU kernel to subtract $\sigma_\ell$ from the diagonal of $Z_{2}^{(\ell )}$, for all $\ell=1, \ldots, s$\;
}
\For{{$k=1,n_{b}+1,2n_{b}+1,\ldots n-m$}}
{
    $n_{b}=\min \{ n_{b}, {n-m-k+1}\}$; {$mn_{b}=\min \{ m, n_{b}\}$; $r=k+n_{b}+m$}\; 
    Copy {$\left[ \begin{array}{c} A^{T}\\ -I \end{array}\right](k:2n,k+m:r-1)$ to $Z_{1}(k:2n,1:n_{b})$}\;
    \For{$\ell=1,2,\ldots ,s$}
    {
        {Copy $Z_{2}^{(\ell )}(k:k+n_{b}-1,1:m)$ to $Z_{block}^{(\ell )}(1:n_{b},1:m)$}\;
        {Copy $Z_{1}(k:k+n_{b}-1,1:n_{b})$ to $Z_{block}^{(\ell )}(1:n_{b},m+1:m+n_{b})$}\;
    }
    \If{$n_{b}>m$}
    {
        {
        Run a GPU kernel to subtract $\sigma_\ell$ from the remaining diagonal elements in $Z_{block}^{(\ell)}$, for all $\ell=1, \ldots, s$\;
        }
    }
    {
    Run a GPU kernel to compute the LQ factorization for $Z_{block}^{(\ell )}$, producing $P^{(\ell )^*}$, and $y_{\ell}(k:k+n_{b}-1)$ stored in $w^{(\ell )}(k:k+n_{b}-1)$, for all $\ell=1\ldots s$\;
    }
    Copy {$P^{(\ell )^*}(1:n_{b}+m,n_{b}+1:n_{b}+m)$} to $\mathbf{{d}P}(:,(\ell -1)m+1:\ell m)$\;

    {
    Perform batched multiplications: for all $\ell =1,\ldots, s$ in parallel compute
    }
    \tab {$\mathbf{{d}W}(1:n_{b}+m,\ell)=P^{(\ell )^*}(1:n_{b}+m,1:n_{b})\cdot w^{(\ell )}(k:k+n_{b}-1)$}\;
    \tab {$w^{(\ell )}(k+n_{b}:2n)=w^{(\ell )}(k+n_{b}:2n)-Z_{2}^{(\ell)}(k+n_{b}:2n,1:m)\cdot \mathbf{{d}W}(1:m,\ell)$}\;
    \tab {$Z_{2}^{(\ell )}(k+n_{b}:2n,1:m)=Z_{2}^{(\ell )}(k+n_{b}:2n,1:m)\cdot \mathbf{{d}P}(1:m,(\ell -1)m+1:\ell m)$}\;
    {
    Run a GPU kernel to update $w^{(\ell )}$ and $Z_{2}^{(\ell )}$, for all $\ell=1, \ldots, s$:
    }
    \tab {$w^{(\ell )}(r\!-\!mn_{b}:r\!-\!1)=w^{(\ell )}(r\!-mn_{b}:r\!-1)\!+\!\sigma (\ell )\!\cdot \!\mathbf{{d}W}(n_{b}\!+\!m\!-\!mn_{b}\!+\!1:n_{b}\!+\!m,\ell)$}\;
    \tab {$Z_{2}^{(\ell )}(r-mn_{b}:r-1,1:m)=Z_{2}^{(\ell )}(r-mn_{b}:r-1,1:m)-$ \tab \hspace*{2em}$\sigma (\ell )\cdot\mathbf{{d}P}(m+n_{b}-mn_{b}+1:m+n_{b},(\ell -1)m+1:\ell m)$}\;
}
{$\mathbf{{d}Z_{2}}(k\!+\!n_{b}:2n,1:sm) \!=\! \mathbf{{d}Z_{2}}(k\!+\!n_{b}:2n,1:sm)\!+\!\mathbf{{d}Z_{1}}(k\!+\!n_{b}:2n,1:n_{b})\cdot \mathbf{{d}P}(m\!+\!1:m\!+\!n_{b},1:sm) $}\;
{$\mathbf{{d}w}(k+n_{b}:2n,1:s)=\mathbf{{d}w}(k+n_{b}:2n,1:s)-\mathbf{{d}Z_{1}}(k+n_{b}:2n,1:n_{b})\cdot \mathbf{{d}W}(m+1:m+n_{b},1:s)$}\;
\caption{Parallel GPU/CPU algorithm for computing solution of {$(A-\sigma_{\ell}I)^{T}x=\cc_{\ell}$}, $\ell =1,\ldots,s$.}\label{alg-parallel-trans-diff-rhs}
\end{algorithm}

\setcounter{algocf}{10}

\begin{algorithm}[h!]
\setcounter{AlgoLine}{28}
{
Copy $\mathbf{dZ_{2}}(n-m+1:2n,1:s\cdot m)$ and $\mathbf{dw}(n-m+1:2n,1:s)$ from the GPU to $\mathbf{Z_{2}}(n-m+1:2n,1:s\cdot m)$ and $\mathbf{w}(n-m+1:2n,1:s)$ on the CPU\;
Revert the notation: $Z_{2}^{(\ell)} \equiv \mathbf{Z_{2}}(:,(\ell -1)m+1:\ell m)$ and $w^{(\ell)} \equiv \mathbf{w}(:,\ell)$\;

\vspace*{0.3cm}

\tcp{The following is executed on the CPU.}
}
\For{$\ell =1,2,\ldots ,s$}
{
    \For{$k =1,2,\ldots ,m-1$}
    {
        {Annihilate elements except the first in $Z_{2}^{(\ell )}(r,k:m)$, where $r=n-m+k$}\;
        {Update $Z_{2}^{(\ell )}(r+1:2n,k:m)$}\;
        {$w^{(\ell )}(r)=w^{(\ell )}(r)/Z_{2}^{(\ell )}(r+1,k)$}\;
        {$w^{(\ell )}(r+1:2n)=w^{(\ell )}(r+1:2n)-Z_{2}^{(\ell )}(r+1:2n,k)\cdot w^{(\ell )}(r)$}\;
    }
    {$w^{(\ell )}(n)=w^{(\ell )}(n)/Z_{2}^{(\ell )}(n,m)$}\;
    {$w^{(\ell )}(n+1:2n)=w^{(\ell )}(n+1:2n)-Z_{2}^{(\ell )}(n+1:2n,m)w^{(\ell )}(n)$}\;
    $\mathbf{x}(1:n,\ell )=w^{(\ell )}(n+1:2n)$\;
}
\caption{Parallel GPU/CPU algorithm for computing solution of {$(A-\sigma_{\ell}I)^{T}x=\cc_{\ell}$}, $\ell =1,\ldots,s$ --- continuation.}
\end{algorithm}


\section{Numerical experiments}
\label{nump}

In this section, we demonstrate the effectiveness of the proposed algorithms by running a series of numerical experiments, and
by comparing the obtained timings to the testing results from \cite{BosBD13}.
The following computational environment was used:
\begin{itemize}
	\item 2x Intel(R) Xeon(R) E5-2690 v3 @ 2.60GHz (24 cores in total);
	\item 256 GB RAM, each processor is equipped with 30 MB of cache memory;
	\item Nvidia Tesla K40c (Kepler generation, 12 GB of GDDR5);
	\item Intel Parallel Studio XE 2016 + MKL 11.3;
	\item Nvidia CUDA 8.0.
\end{itemize}

The CPUs reach the peak DGEMM performance of about $800$ Gflops, while the GPU has the peak performance of about $1200$ Gflops. This represents our main computing machine.

We also performed the same experiments on an older machine with 2x Intel(R) Xeon(R) CPU E5620 (2.40GHz, 8 cores in total) and Nvidia Tesla S2050 Computing System (Fermi generation). On this machine CPUs reach the peak DGEMM performance of about $70$ Gflops, while the GPU has the peak performance of about $300$ Gflops. The results in this case showed that the GPU-bound algorithms obtained better improvement when compared to the CPU-bound versions.
\\ \

\subsection{$m$-Hessenberg reduction}
\begin{figure}
	\begin{minipage}{1\textwidth}
	\centering

    \subfloat[Ratios of execution times between the CPU-bound Algorithm \ref{mhess:alg:outer} and the hybrid Algorithm \ref{mhess:alg:GPU_outer}.]
        {\includegraphics[width=.48\textwidth]{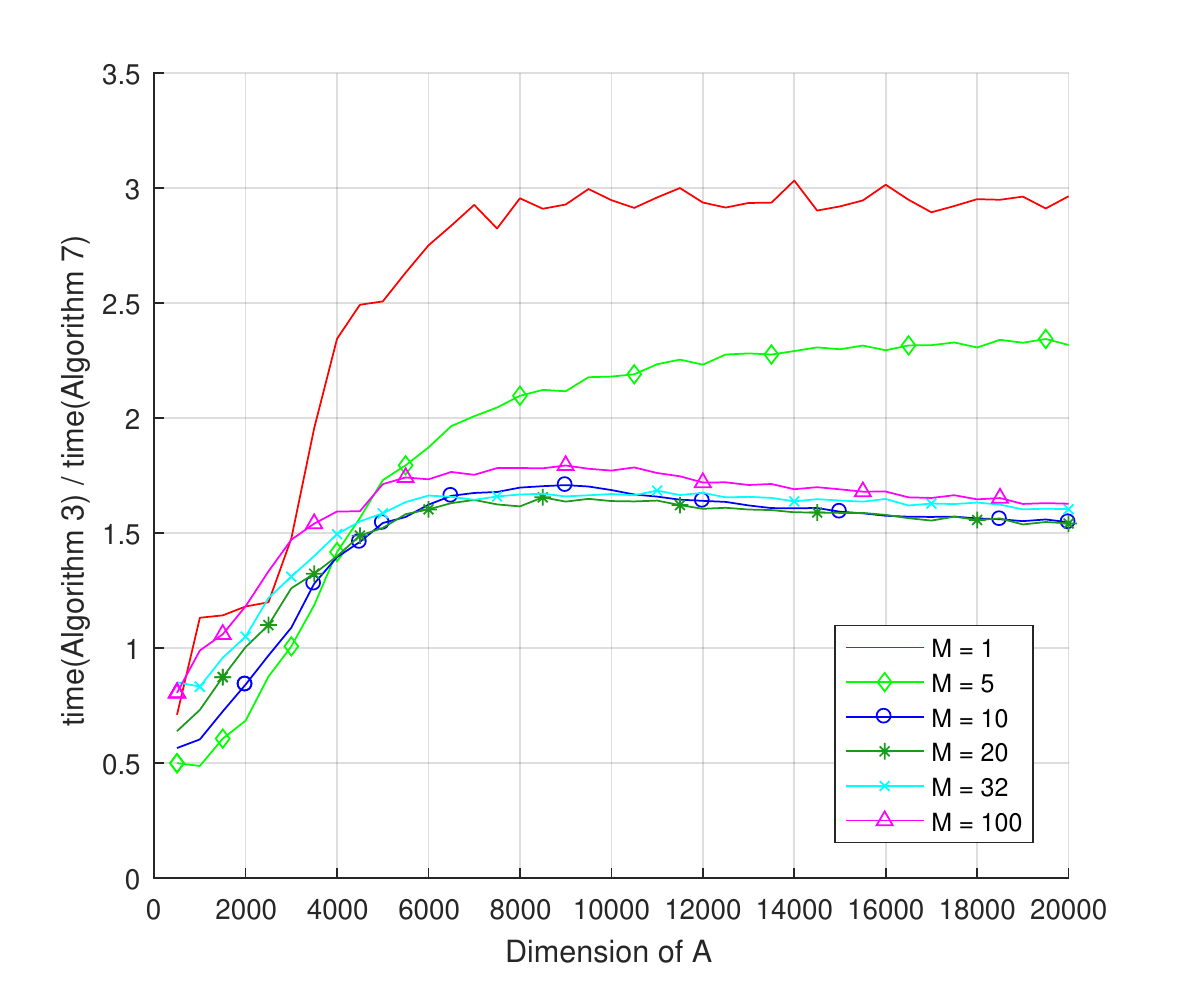}
        \label{num:fig:hess_speedup_pascal}}
    \hfill
    \subfloat[The Gflops rates for the hybrid Algorithm \ref{mhess:alg:GPU_outer}.]
        {\includegraphics[width=.48\textwidth]{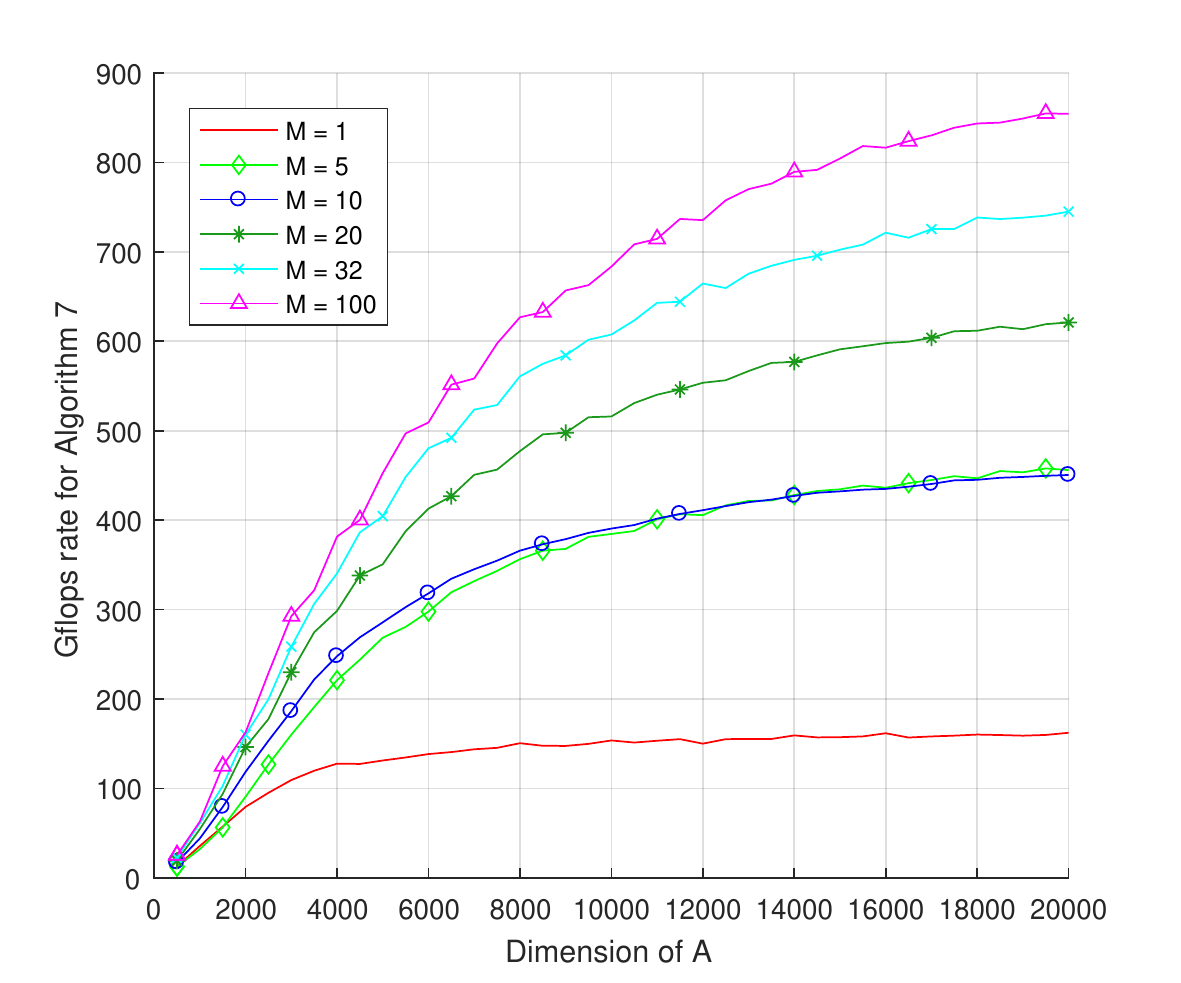}
        \label{num:fig:hess_gflops_pascal}}
    \\
    \end{minipage}
    \caption{\label{num:fig:hess_pascal} Performance of the hybrid $m$-Hessenberg reduction algorithm.}
\end{figure}

In the first experiment, we run the algorithms for reducing the matrix $A\in\Real^{n \times n}$ to
the $m$-Hessenberg form, for increasing values of $n$ and various $m$.
On our computing platform, the best performance of the hybrid algorithm is obtained by setting the value of the block size $b$ to $128$ or $192$,
depending on $m$, and by using $12$ CPU cores for block processing, with the remaining $12$ cores doing the CPU update part of
Algorithm \ref{mhess:alg:GPU_outer}, as described in Section \ref{mhess}.
Figure \ref{num:fig:hess_speedup_pascal} shows the speedup this algorithm obtains when running against the CPU-bound
Algorithm \ref{mhess:alg:outer} from \cite{BosBD13} on our main machine. As we can see, the speedup factors spread from 1.5 up to 3.
The hybrid algorithm reaches the performance of up to $800$ Gflops for larger values of $m$.

On the other hand on our older machine parallel algorithm outperforms its CPU-bound counterparts by the factors going from 2.3 up to 4.8 for larger dimensions $n$, and achieving at most 165 Gflops.

The overall performance of the algorithm is clearly better for larger $m$ since the updates from the right in the block processing phase
are less frequent, and block sizes during the updates are larger and result in better performance of the GEMM-like routines.
The CPU-bound algorithm benefits from the same effects, so the relation of the speedup to the value of $m$ is not so immediate.

The number of operations for the $m$-Hessenberg reduction is $10/3n^3 + 5/2n^2 b - 9/2 n^2 m + n^2 m^2/(2b)$, as stated in \cite{BosBD13}.

\

\subsection{Transfer function evaluation}\label{s:shifted-sys-results}
\begin{figure}
	\begin{minipage}{1\textwidth}
	\centering

    \subfloat[Ratios of execution times between the CPU-bound and the GPU-bound algorithms for solving shifted systems.]
        {\includegraphics[width=.48\textwidth]{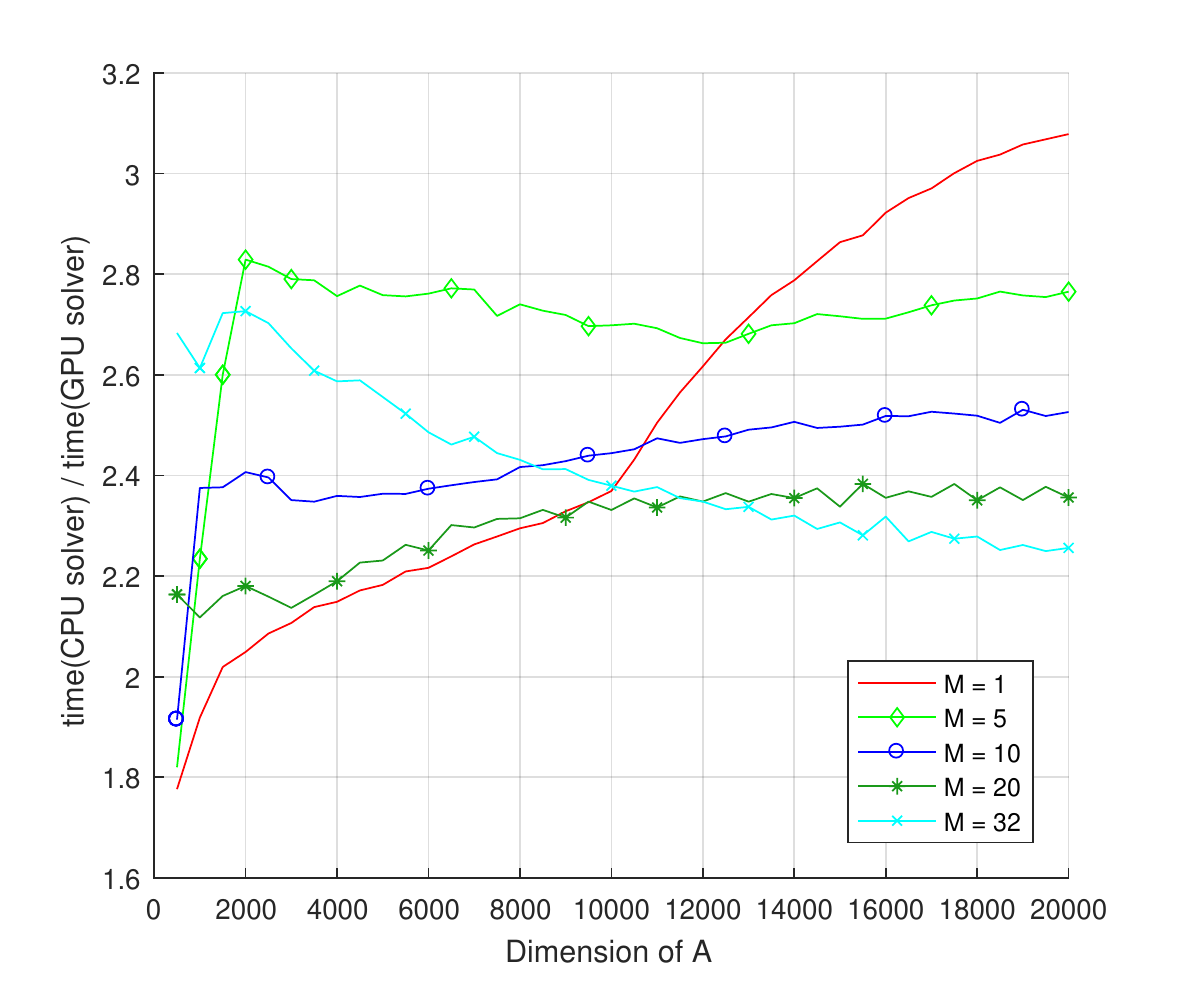}
        \label{num:fig:freq_speedup_pascal}}
    \hfill
    \subfloat[The Gflops rates for the GPU-bound algorithm.]
        {\includegraphics[width=.48\textwidth]{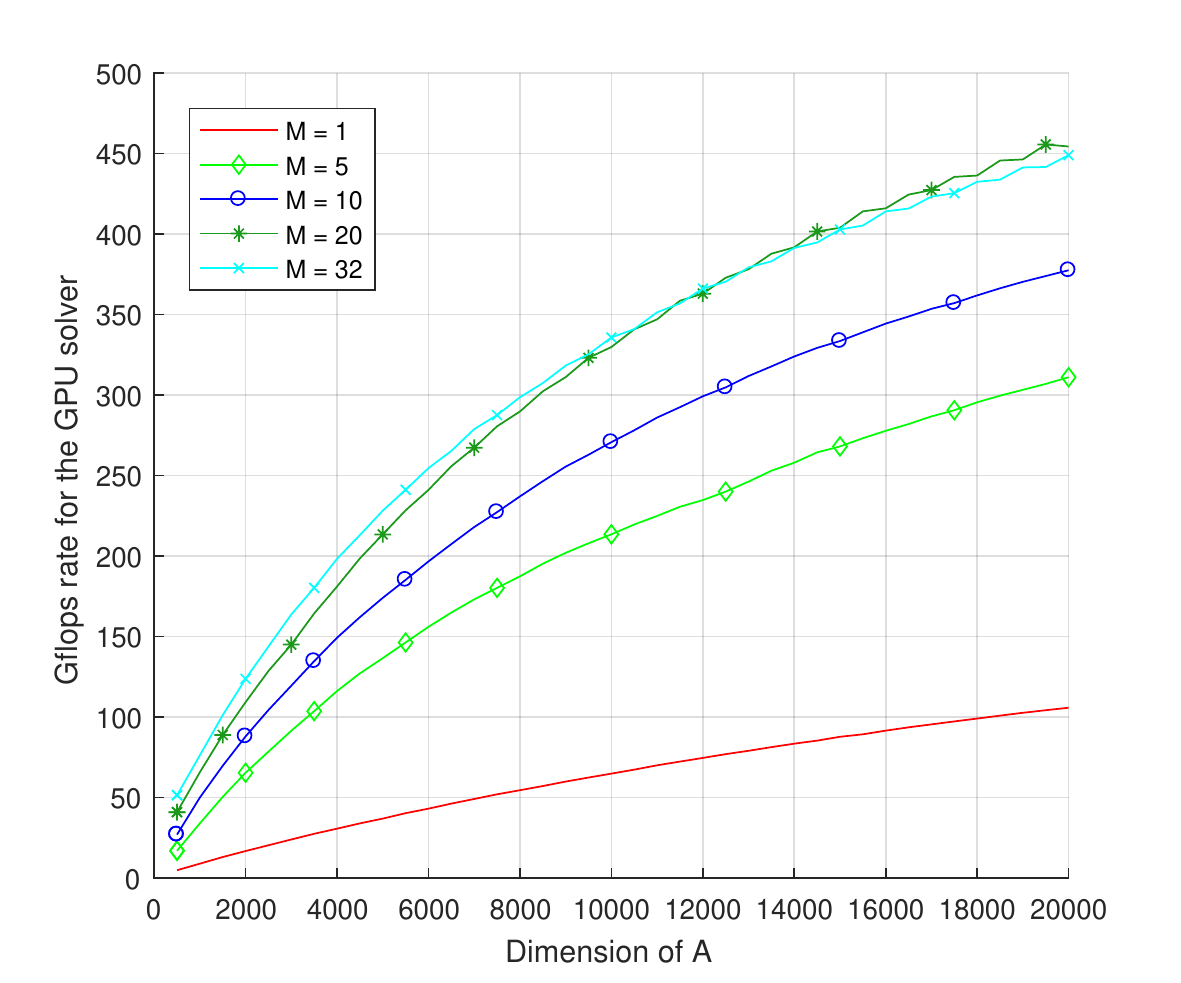}
        \label{num:fig:freq_gflops_pascal}}
    \\
    \end{minipage}
    \caption{\label{num:fig:freq_pascal} Performance of the GPU-bound algorithm for solving shifted systems.}
\end{figure}

Next, we compared  the algorithms for transfer function evaluations using
random systems ($A$, $B$, $C$) of dimensions up to $n=20000$.
Each generated system is converted to the controller-Hessenberg form.
We then measured the time needed evaluate $G(\sigma)=C(\sigma I - A)^{-1}B$ at $1000$ random values of $\sigma$, using the CPU and the GPU variants
of the algorithm described in Section \ref{freq}.
Each algorithm was run with the block size set to the optimal value of $64$, $96$ or $128$ (depending on $m$), and each batch contained $200$ shifts.
The results are shown in Figure \ref{num:fig:freq_pascal}.
Once again, we obtain a decent speedup compared to the CPU-bound routine---depending on $m$, the GPU algorithm is between $2.3$ and $2.8$
times faster, which is better than the DGEMM flop ratio of the two devices. Thus, this algorithm is better suited for GPUs than for CPUs.

The speed up factors obtained on the older machine are going from 2.9 up to 4.2 for larger dimensions $n$.

The number of operations per shift, not including the controller-Hessenberg reduction, is
$
    4n^2 m + 4 n (n-m) m^2/n_b + 4 n n_b m + 40 n m^2 + 8n m p
$,
as stated in \cite{BosBD13}.

\

\subsection{Simultaneous RQ-factorizations on the GPU}
\begin{figure}
    \begin{minipage}{1\textwidth}
    \centering

    \subfloat[The Gflops rates for the RQ-kernel on the GPU.]
        {\includegraphics[width=.48\textwidth]{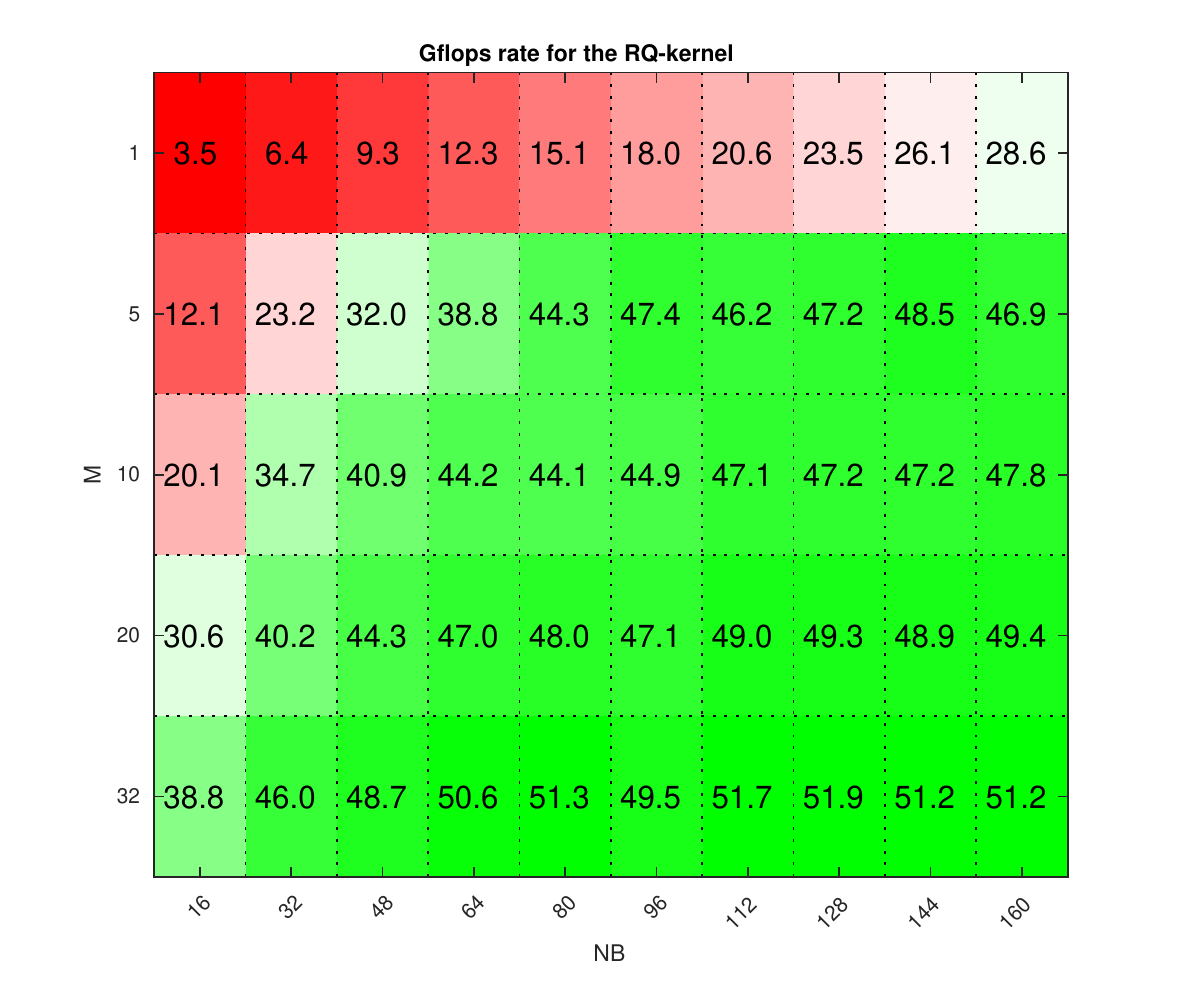}
        \label{num:fig:rqGflop_pascal}}
    \hfill
    \subfloat[The percentage of the total execution time within the system solver taken by the RQ-kernel.]
        {\includegraphics[width=.48\textwidth]{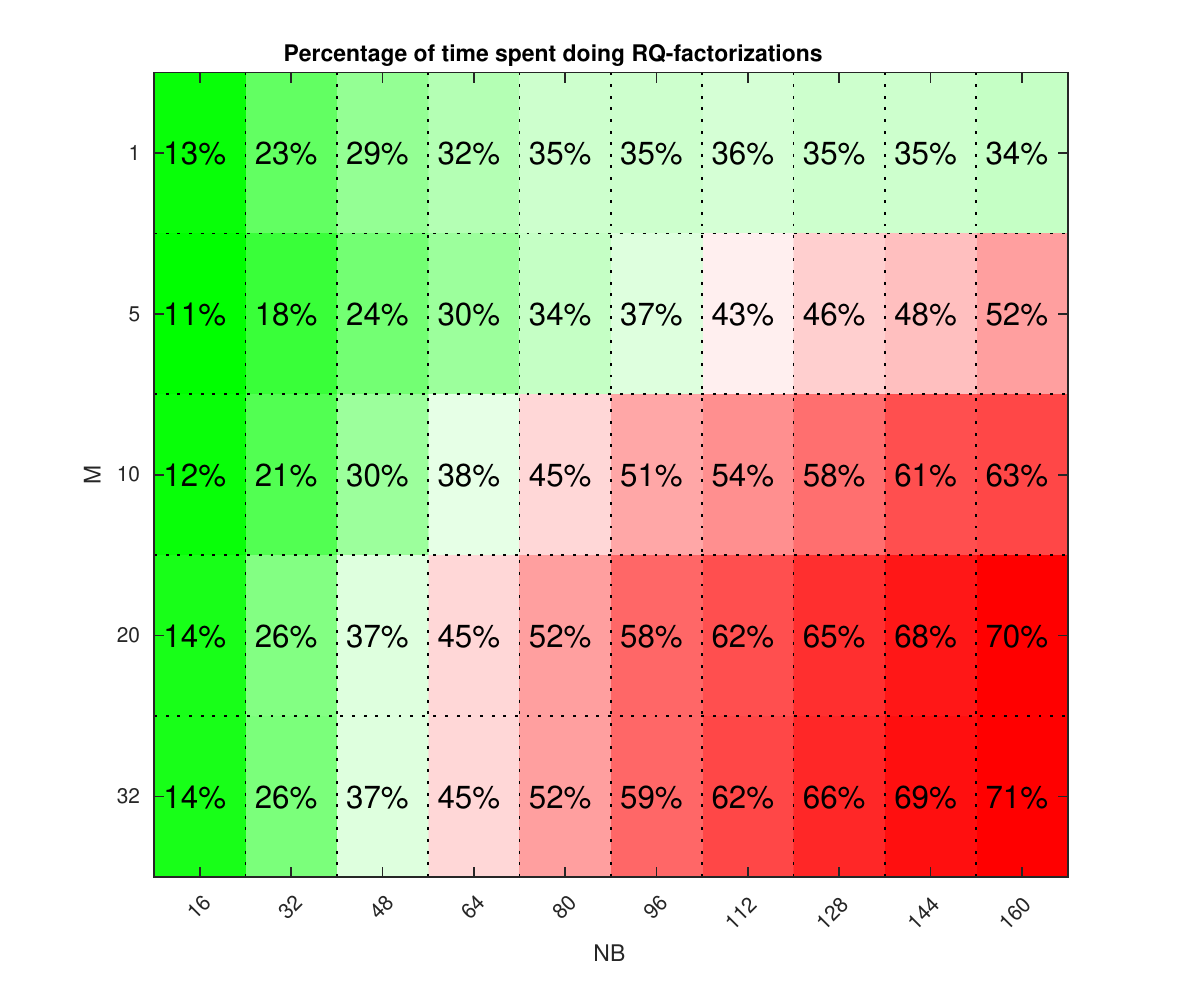}
        \label{num:fig:rqPercentage_pascal}}
    \\
    \end{minipage}
    \caption{\label{num:fig:rq_pascal} Performance of the kernel computing simultaneous RQ-factorizations on the GPU. (Total time here
        does not include the controller-Hessenberg reduction.)}
\end{figure}

One obvious bottleneck of the GPU-bound system solver is Algorithm \ref{alg-parallel-RQ} --
computation of the RQ-factorizations of the block matrices $Z_{block}^{(\ell )}$, simultaneously for all shifts in the same batch.
Since all these (complex) matrices are of small dimension $n_b \times (n_b + m)$, the benefit of parallelism in the arithmetic
is heavily overshadowed by the memory traffic needed to fetch and store elements of these matrices from the global GPU memory into the registers.
Thus the performance of this part of the system solver is limited by the global memory bandwidth of the GPU.
Figure \ref{num:fig:rq_pascal} shows the Gflops rate for the simultaneous RQ-factorizations during one run of the system solver test,
as described in the previous paragraph, and the percentage of time spent doing RQ-factorization during the entire
system solve phase. The larger block-size is very beneficial for fast BLAS3 routines on the GPU, and clearly,
the non-blocked RQ-kernel becomes a bottleneck, preventing the Gflops rate from going even higher.
For the Tesla K40c card, the theoretical memory bandwidth is 288GB/s, and since the RQ-kernel has 2.5 double precision flops per one
load or store from the global memory, the maximum theoretical rate for that algorithm is 92.5 Gflops.
The number of real double precision operations per single RQ-factorization of a double complex trapezoidal $n_b \times (n_b + m)$ matrix is
$m n_b (30 n_b + 20m + 37)$.

Note that {cuBLAS}, starting from the version 6.5, also contains a routine \alg{cublasZgeqrfBatched} for computing a batch of QR-factorizations
simultaneously. This routine, which in contrast to ours uses Householder reflectors, achieves similar performance levels.
However, for now {cuBLAS} does not contain a routine for applying a batch of reflectors.

\

\subsection{The combined performance for solving shifted systems}
Different parts of the algorithm for solving shifted systems presented in this paper have peak performance for different combinations of parameters.
The following table shows the time distribution {in case of transfer function evaluation} for various choices of $m$; we fixed the $n=15000$, and solved the systems for $1000$ shifts.

\begin{center}
    \begin{tabular}{|c||c|c|c|}
        \hline
        $m$                  & $1$             & $5$              & $32$ \\ \hline\hline
        contr-Hess reduction & $88.5\%$        & $60.8\%$         & $25.6\%$ \\ \hline
        \multirow{2}{*}{small batched RQ}
                             & $3.0\%$         & $10.0\%$         & $32.6\%$ \\
                             & $23.51$ Gflops  & $47.44$ Gflops   & $51.87$ Gflops \\ \hline
        \multirow{2}{*}{batched GEMM}
                             & $1.7\%$         & $8.7\%$          & $12.7\%$ \\
                             & $4.87$ Gflops   & $55.74$ Gflops   & $440.11$ Gflops \\ \hline
        \multirow{2}{*}{outer GEMM}
                             & $1.3\%$         & $9.2\%$          &  $20.2\%$ \\
                             & $834.12$ Gflops & $1012.27$ Gflops & $1108.52$ Gflops \\ \hline \hline
        total time           & $83.10$s        & $48.0$s          & $128.25$s \\ \hline
    \end{tabular}
\end{center}

For smaller $m$, almost the entire time is spent performing the Hessenberg reduction, while the distribution is more balanced for larger
$m$. The outer GEMM operation multiplies large matrices regardless of $m$, and always attains the peak performance. On the other hand,
the batched GEMM multiplies small matrices when $m$ is small, which is not well suited for such routine. Its performance increases by a
large margin for larger $m$.

{
\subsection{Performance of linear solvers in IRKA}\label{sec:irka:results}
As mentioned before, the IRKA algorithm for model order reduction is heavily burdened by a large number of shifted linear systems that need to be solved in every iteration. To show the benefits of using the algorithms presented in this paper, we implemented two versions of the IRKA algorithm. In both versions we embedded our framework for solving shifted systems: the initial reduction to the controller Hessenberg form, and the algorithms described in \S \ref{subs:irka:syst}. In the first version, both of these components were implemented on the CPU only, while the second version used the hybrid CPU+GPU implementation. Other parts of the IRKA algorithm were implemented on the CPU, as they deal with matrices of small dimensions. The CPU implementation is based on the LAPACK \cite{LAPACK} library, provided in the Intel's MKL.

Since this work focuses on the efficiency of running IRKA iterations, and not on the issues concerning the convergence of the IRKA algorithm, we fixed both algorithms to always run $30$ iterations.
The integral implementation of IRKA is a separate ongoing work beyond the scope of this paper.

We generated the test LTI systems at random; the dimension $n$ of the matrix $A$ was between $3000$ and $15000$, and we studied three choices of parameters $m$ and $p$: $m=p=1$, $m=p=10$, and $m=p=32$. In all cases we ran the $30$ IRKA iterations with the goal of reducing the system to dimension $r=100$. The results are shown in Figure \ref{num:fig:irka_pascal}.

\begin{figure}
	\begin{minipage}{1\textwidth}
	\centering

    \subfloat[Comparison of the execution times of two IRKA algorithms: one incorporating the CPU-bound and the other using the hybrid algorithm.]
        {\includegraphics[width=.48\textwidth]{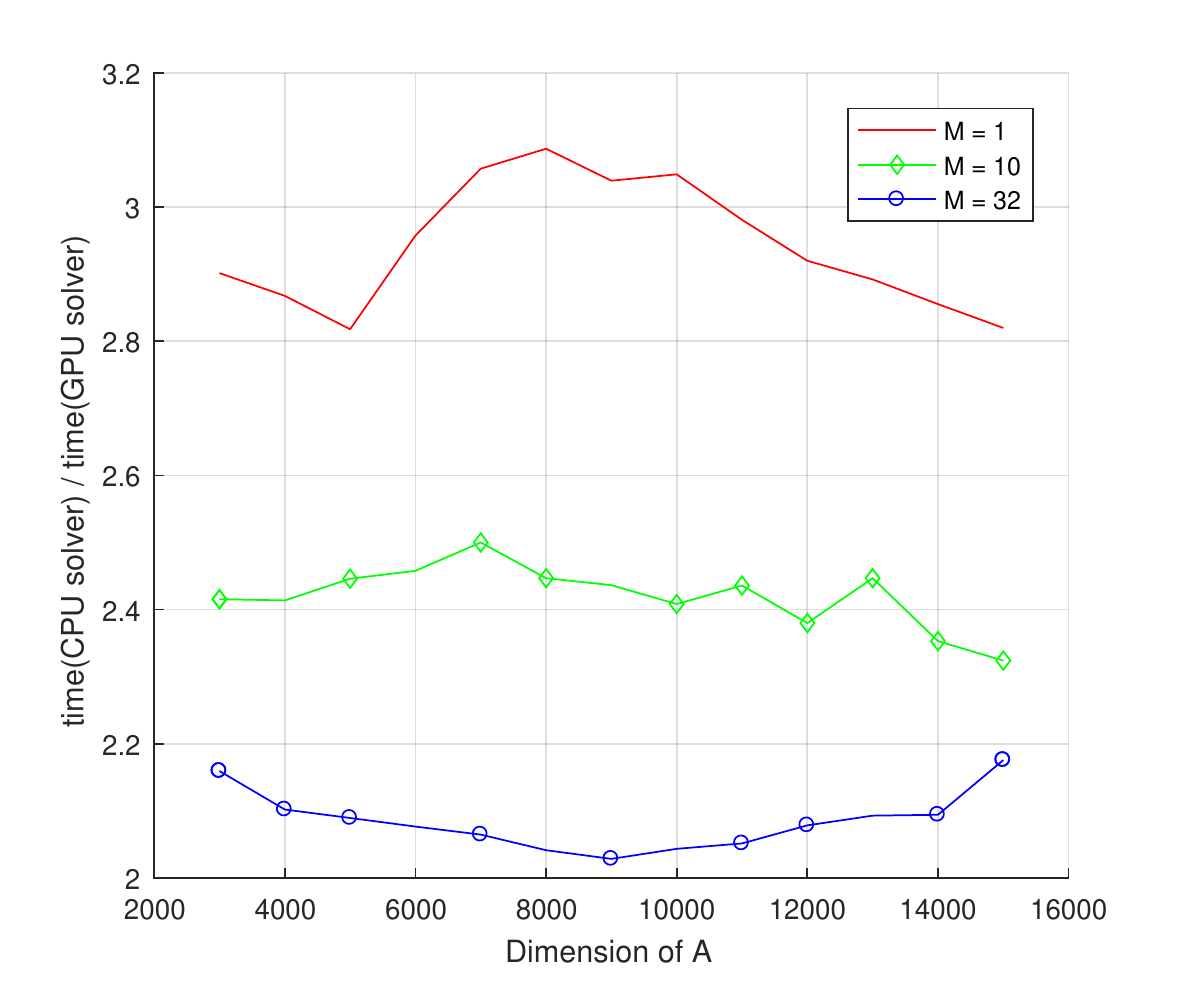}
        \label{num:fig:irka_speedup_pascal}}
    \hfill
    \subfloat[Comparison of the times spent on solving shifted systems in the two IRKA algorithms.]
        {\includegraphics[width=.48\textwidth]{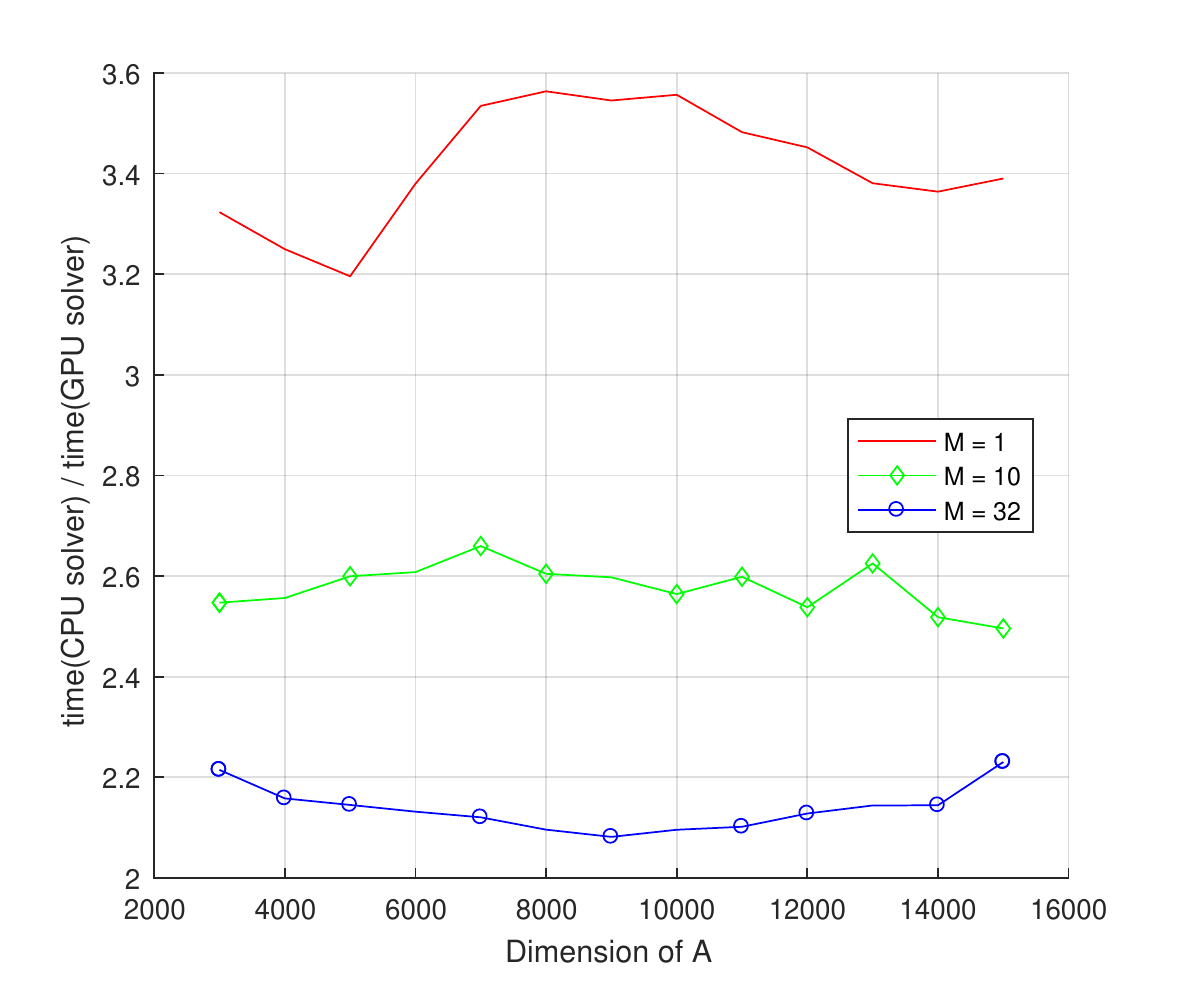}
        \label{num:fig:irka_solvers_speedup_pascal}}
    \\
    \end{minipage}
    \caption{\label{num:fig:irka_pascal} Performances of the CPU-bound and the hybrid IRKA algorithm.}
\end{figure}

As before, we obtain a decent speedup compared to the CPU-bound routine. Depending on $m$, the hybrid algorithm is from $2$ up to $3.1$ times faster. When we compare only the time spent on solving the shifted systems within IRKA, the speedup factors are even higher, as shown in Figure \ref{num:fig:irka_solvers_speedup_pascal}, ranging from $2.1$ up to $3.6$. Again, it is interesting to examine the time distribution for various choices of $m$ and fixed $n=15000$, which is shown in the following table for the hybrid algorithm.

\begin{center}
    \begin{tabular}{|c||c|c|c|}
        \hline
        $m=p$                                         & $1$        & $10$       & $32$ \\
        \hline
        \hline
        contr-Hess reduction                          & $42.63\%$  & $10.05\%$  & $3.36\%$ \\
        \hline
        solver for $(A-\sigma_{\ell}I)^{T}x=\cc_{\ell}$ & $30.88\%$  & $48.67\%$  & $52.46\%$ \\
        \hline
        solver for $(A-\sigma_{\ell}I)x=\bb_{\ell}$     & $22.20\%$  & $38.36\%$  & $42.39\%$\\
        \hline
        \hline
        total time                                    & $171.87$s  & $305.92$s  & $900.55$s\\
        \hline
    \end{tabular}
\end{center}


Here the portion of time spent on performing the controller Hessenberg reduction again decreases as $m$ increases, but for $m=1$ it takes almost a half of the total execution time. For larger $m$ the time spent on solving shifted systems becomes dominant. In addition, the solver for solving transposed shifted systems $(A-\sigma_{\ell}I)^{T}x=\cc_{\ell}$ is more expensive than the solver for $(A-\sigma_{\ell}I)x=\bb_{\ell}$, since its right-hand side is unreduced. These results illustrate the importance of the simultaneous reduction of both the system matrix and the right-hand side, where the right hand side has only a small number of nontrivial components.
} 
\section{Conclusion}
In this paper we propose a combination of a hybrid CPU-GPU and a pure GPU algorithm for solving shifted linear systems of the form $(A - \sigma I)X = B$, for a large number of shifts $\sigma\in\mathbb{C}$ and multiple right-hand sides. This is done in two phases: The first phase reduces the pair $(A,B)$ to the controller Hessenberg form, and in the second phase the shifted systems in $m$-Hessenberg form are solved by processing the shifts in batches. For each batch of shifts, the corresponding $m$--Hesseberg systems are reduced to triangular forms by simultaneous RQ factorizations. 
The reduction in the first phase is implemented as a highly parallel CPU-GPU hybrid algorithm, and the solver in the second phase is implemented entirely on the GPU.  The benefits of such a load distribution are demonstrated by numerical experiments. We provided detailed blueprints that can be used for case study and further development.\\

\noindent We believe that the proposed algorithm with its software implementation will prove to be useful tool in a variety of applications, where it can be used as one of the core computational routines. Its potential is illustrated in an efficient implementation of transfer function evaluation and in the model order reduction algorithm IRKA \cite{IRKA}. The latter will be pursued toward an efficient model reduction software toolbox. 

\section{Acknowledgements} We wish to thank  Vedran Novakovi\'{c} (STFC Daresbury Laboratory) for providing insights and sharing his expertise that helped this work.

\nocite{IRKA}
\nocite{Beattie}
\nocite{LaubLinnemann}
\bibliographystyle{plain}
\bibliography{5_Bibliography}

\end{document}